\documentclass[journal]{IEEEtran}

\usepackage{cite}
\usepackage{amssymb}
\usepackage{amsmath}
\usepackage{graphicx}
\usepackage{sidecap}
\usepackage{subcaption}
\usepackage[]{algorithm2e}

\usepackage{enumitem}
\usepackage{cleveref}
\usepackage{multirow}
\usepackage{tabularx}
\usepackage{epstopdf}
\usepackage{float}
\usepackage{hhline}
\usepackage[symbol]{footmisc}

\usepackage{color,soul}
\soulregister\cite7
\soulregister\ref7

\usepackage{xcolor}

\makeatletter

\newcommand{\Rmnum}[1]{\expandafter\@slowromancap\romannumeral #1@}

\makeatother

\begin{document}
%
\title{LIDIA: Lightweight Learned Image Denoising \\ with Instance Adaptation}

\author{Gregory~Vaksman,
        Michael~Elad and 
        Peyman Milanfar
\thanks{\noindent G. Vaksman is with the Department of Computer Science, Technion Institute of Technology, Technion City, Haifa 32000, Israel,
[grishav@campus.technion.ac.il].
M. Elad and P. Milanfar are with Google Research, Mountain-View, California [melad/milanfar@google.com].}
}

\maketitle

\graphicspath{{./figures/schemes_jpg/}{./figures/pictures/denoising_jpg/}{./figures/pictures/adaptation_jpg/}}

\begin{abstract}

Image denoising is a well studied problem with an extensive activity that has spread over several decades. Despite the many available denoising algorithms, the quest for simple, powerful and fast denoisers is still an active and vibrant topic of research. Leading classical denoising methods are typically designed to exploit the inner structure in images by modeling local overlapping patches, while operating in an unsupervised fashion. In contrast, recent newcomers to this arena are supervised and universal neural-network-based methods that bypass this modeling altogether, targeting the inference goal directly and globally, while tending to be very deep and parameter heavy. 
	
This work proposes a novel lightweight learnable architecture for image denoising, and presents a combination of supervised and unsupervised training of it, the first aiming for a universal denoiser and the second for adapting it to the incoming image. 
Our architecture embeds in it several of the main concepts taken from classical methods, relying on patch processing, leveraging non-local self-similarity, exploiting representation sparsity and providing a multiscale treatment. Our proposed universal denoiser achieves near state-of-the-art results, while using a small fraction of the typical number of parameters. In addition, we introduce and demonstrate two highly effective ways for further boosting the denoising performance, by adapting this universal network to the input image.  

\end{abstract}


\section{Introduction}
	
Image denoising is a well studied problem, and many successful algorithms have been developed for handling this task over the years, e.g.  NLM~\cite{Buades_Non_Local_Means_2005}, KSVD~\cite{Elad_Image_Denoising_KSVD_2006}, BM3D~\cite{Egiazarian_BM3D_2007}, EPLL~\cite{Zoran_Weiss_EPLL_2011}, WNNM~\cite{WNNM_2014} and others~\cite{roth2009,Sapiro_LSSC_2009,Sapiro_PLE_2012,dong2012nonlocally,Ram_Patch_Ordering_2013,Milanfar_Tour_of_Modern_Image_Filtering_2013,Milanfar_General_Framework_for_Image_Restoration_2014,romano2015boosting,Vaksman_Patch_Ordering_2016,Yair_2018_CVPR,Sulam_Multi_Scale_Dictionaries_2014,Papyan_Multi_Scale_EPLL_2015}. These classically-oriented algorithms strongly rely on models that exploit properties of natural images, usually employed while operating on small fully overlapped patches. For example, both EPLL~\cite{Zoran_Weiss_EPLL_2011} and PLE~\cite{Sapiro_PLE_2012} perform denoising using Gaussian Mixture Modeling (GMM) imposed on the image patches. The K-SVD algorithm~\cite{Elad_Image_Denoising_KSVD_2006} restores images using a sparse modeling of such patches.  BM3D~\cite{Egiazarian_BM3D_2007} exploits self-similarity by grouping similar patches to 3D blocks and filtering them jointly. The algorithms reported in~\cite{Sulam_Multi_Scale_Dictionaries_2014,Papyan_Multi_Scale_EPLL_2015} harness a multi-scale analysis framework on top of the above-mentioned local models. Common to all these and many other classical denoising methods is the fact that they operate in an unsupervised fashion, adapting their treatment to each image. 

Recently, supervised deep-learning based methods entered the denoising arena, showing state-of-the-art (SOTA) results in various contexts \cite{Burger_MLP_Denoising_2012,Chen_TNRD_2015,wang2015deep,Zhang_DnCnn_2017,Mao_RED_2016,Zhang_FFDNet_2018,remez2018class,Tai_MemNet2017,Liu_MwCnn2018,Zhang_RDN_2018,Liu_NLRN_2018}. In contrast to the above-mentioned classical algorithms, deep-learning based methods tend to bypass the need for an explicit modeling of image redundancies, operating instead by directly learning the inference from the incoming images to their desired outputs. In order to obtain a non-local flavor in their treatment, as self-similarity or multi-scale methods would do, most of these algorithms (\cite{Liu_NLRN_2018} being an exception) tend to increase their footprint by utilizing very deep and parameter heavy networks. These reflect badly on their memory consumption, the required amount of training images and the time for training and inference. Note that most deep-learning based denoisers operate in an universal fashion, i.e., they apply the same trained network to all incoming images.

	\begin{figure}[t]
	    \centering
		\begin{subfigure}{0.24\textwidth}
    	    \captionsetup{justification=centering}
			\includegraphics[width=\textwidth]{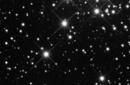}
			\caption{Clean image}
			\label{fig:demo:clean_astronomical}
			\vspace*{6pt}
		\end{subfigure}
		\begin{subfigure}{0.24\textwidth}
    	    \captionsetup{justification=centering}
			\includegraphics[width=\textwidth]{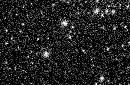}
			\caption{Noisy (${\sigma = 50}$)}
			\label{fig:demo:noisy_astronomical}
			\vspace*{6pt}
		\end{subfigure}
		\begin{subfigure}{0.24\textwidth}
    	    \captionsetup{justification=centering}
			\includegraphics[width=\textwidth]{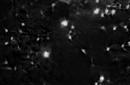}
			\caption{Denoised (before adaptation) \\ PSNR = 24.33dB}
			\label{fig:demo:nlms_d1_astronomical}
			\vspace*{6pt}
		\end{subfigure}
		\begin{subfigure}{0.24\textwidth}
    	    \captionsetup{justification=centering}
			\includegraphics[width=\textwidth]{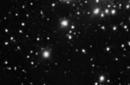}
			\caption{Denoised (after adaptation) \\ PSNR = 26.25dB}
			\label{fig:demo:nlms_d2_astronomical}
			\vspace*{6pt}
		\end{subfigure}
		\caption{Example of our adaptation approach: (a)~and~(b) show the clean and noisy images respectively; (c) is the denoised result by our universal architecture trained on 432 BSD500 images~\cite{martin2001_bsd_database}; (d) presents our externally adapted result, using an additional single astronomical image (shown in Figure~\ref{fig:train_astronomical}).}
		\label{fig:demo}
	\end{figure}

An interesting recent line of work by Lefkimmiatis proposes a denoising network with a significantly reduced number of parameters, while persisting on near SOTA performance~\cite{Lefkimmiatis_NLNet_2017,Lefkimmiatis_UNLNet_2018}. This method leverages the non-local self-similarity property of images by jointly operating on groups of similar patches. The network's architecture consists of several repeated stages, each resembling a single step of the proximal gradient descent method under sparse modeling~\cite{parikh2014proximal}. In comparison with DnCNN~\cite{Zhang_DnCnn_2017}, the work reported in~\cite{Lefkimmiatis_NLNet_2017,Lefkimmiatis_UNLNet_2018} shows a reduction by factor of $14$ in the number of parameters, while achieving denoising PSNR that is only $\sim0.2$~dB lower.

In this paper we propose two threads of novelty. Our first contribution aims at a better design of a denoising network, inspired by Lefkimmiatis' work. This network is to be trained in a supervised fashion, creating an effective universal denoiser for all images. Our second contribution extends the above by introducing image adaptation: We offer ways for updating the above network for each incoming image, so as to accommodate better its content and inner structure, leading to better denoising.  

Referring to the first part of this work (our universal denoiser), we continue with Lefkimmiatis' line of lightweight networks and propose a novel, easy, and learnable architecture that harnesses several main concepts from classical methods: (i) Operating on small fully overlapping patches; (ii) Exploiting non-local self-similarity; (iii) Leveraging representation sparsity; and (iv) Employing a multi-scale treatment. Our network resembles the one proposed in~\cite{Lefkimmiatis_NLNet_2017,Lefkimmiatis_UNLNet_2018}, with several important differences:
\begin{itemize}
    \item We introduce a multi-scale treatment in our network that combats spatial artifacts, especially noticeable in smooth regions \cite{Sulam_Multi_Scale_Dictionaries_2014}. While this change does not reflect strongly on the PSNR results, it has a clear visual contribution; 
    \item Our network is more effective by operating in the residual domain, similar to the approach taken by~\cite{Zhang_DnCnn_2017}; 
    \item Our patch fusion operator includes a spatial smoothing, which adds an extra force to our overall filtering; and 
    \item Our architecture is trained end-to-end, whereas  Lefkimiatis's scheme consists of a greedy training of the separate layers, followed by an end-to-end warm-start update.
\end{itemize}

\noindent Our proposed method operates on all the overlapping patches taken from the processed image by augmenting each with its nearest neighbors and filtering these jointly. The patch grouping stage is applied only once before any filtering, and it is not a part of the learnable architecture. Each patch group undergoes a series of trainable steps that aim to predict the noise in the candidate patch, thereby operating in the residual domain. 
As already mentioned above, our scheme includes a multi-scale treatment, in which we fuse the processing of corresponding patches from different scales. 

Moving to the second novelty in this work (image adaptation), we present two ways for updating the universal network for any incoming image, so as to further improve its denoised result. This part relies on three key observations: (i) A universally trained denoiser is necessarily less effective when handling images falling outside the training-set statistics; (ii) While our universal denoiser does exploit self-similarity, this can be further boosted for images with pronounced repetitions; and (iii) As our universal denoiser is lightweight, even a single image example can be used for updating it without overfitting. 

In accordance with these, we propose to retrain the universal network and better tune it for the image being served. One option, the external boosting, suggests taking the universally denoised result, and using it for searching the web for similar photos. Taking even one such related image and performing few epochs of update to the network may improve the overall performance. This approach is extremely successful for images that deviate from the training-set, as illustrated in Figure~\ref{fig:demo}. The second adaptation technique, the internal one, re-trains the network on the denoised image itself (and a noisy version of it). This is found to be quite effective for images with marked inner-similaries. 

\noindent To summarize, this paper has two key contributions: 
\begin{enumerate}
\item We propose a novel architecture that is inspired by classical denoising algorithms.
Employed as a universal denoiser and trained in a supervised fashion, our network gives near-SOTA results while using a very small number of parameters to be tuned\footnote{We emphasize that, while the number of parameters in our proposed network is relatively small, this does not imply a similar reduction in computations. Indeed, our inference computational load is similar to DnCNN's.}. 

\item We present an image-adaptation option, in which the above network is updated for better treating the incoming image. This adaptation becomes highly effective in cases of images deviating from the natural image statistics, or in situations in which the incoming image exhibits stronger inner-structure. In these cases, denoising results are boosted dramatically, surpassing known supervised deep-denoisers.   


\end{enumerate}

	

\section{The Proposed Universal Network}
\label{sec:algorithm}


\subsection{Overall Algorithm Overview}

\begin{figure*}
    \centering
    \includegraphics[width=0.96\textwidth]{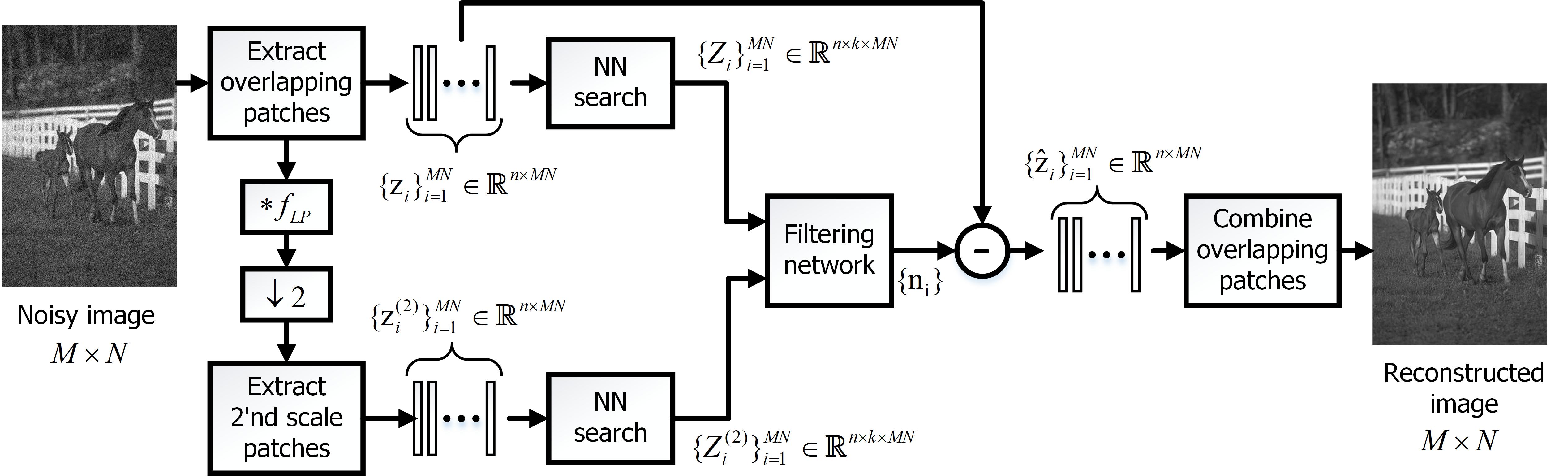}
    \caption{The proposed denoising Algorithm starts by extracting all possible overlapping patches and their corresponding reduced-scale ones. Each patch is augmented with its $k - 1$ nearest neighbors and filtered, while fusing information from both scales. The reconstructed image is obtained by combining all the filtered patches via averaging.}
    \label{fig:alg}
\end{figure*}

Our proposed method extracts all possible overlapping patches of size $\sqrt{n} \times \sqrt{n}$ from the processed image, and cleans each in a similar way. The final reconstructed image is obtained by combining these restored patches via averaging. The algorithm is shown schematically in Figure~\ref{fig:alg}.

In order to formulate the patch extraction, combination and filtering operations, we introduce some notations. Assume that the processed image is of size $M \times N$. We denote by ${Y, \hat{Y} \in \mathbb{R}^{MN}}$ the noisy and denoised images respectively, both reshaped to a 1D vector. Similarly, the corrupted and restored patches in location $i$ are denoted by $\mathbf{z}_i, \mathbf{\hat{z}}_i \in \mathbb{R}^{n}$ respectively, where ${i = 1, \dots, MN}$.\footnote{Note that we handle boundary pixels by padding the processed image using mirror reflection with $\left \lfloor{\sqrt{n}/2}\right \rfloor$ pixels from each side. Thus, the number of extracted patches is equal to the number of pixels in the image.} $\mathbf{R}_i$ denotes the matrix that extracts a patch centered at the $i$-th location. The patch extraction operation from the noisy image is given by $\mathbf{z}_i = \mathbf{R}_i Y$, and the denoised image is obtained by combining the denoised patches $\left\{\mathbf{\hat{z}}_i\right\}$ using weighted averaging, 
\begin{equation}\label{eq:Agg}
    \hat{Y} = \left(\sum_{i}w_i \mathbf{R}_i^T \mathbf{R}_i\right)^{-1}\sum_{i}w_i \mathbf{R}_i^T\mathbf{\hat{z}}_i ,
\end{equation}
where smooth patches get higher weights. More precisely,
\begin{equation}
    w_i = \exp\left\{-\beta \cdot var\left(\mathbf{\hat{z}}_i\right) \right\} \;,
\end{equation}
where $var\left(\mathbf{z}\right)$ is a sample variance of $\mathbf{z}$, and $\beta$ is learned.


\subsection{Our Scheme: A Closer Look}

Zooming in on the local treatment, it starts by augmenting the patch $\mathbf{z}_i$ with a group of its $k-1$ nearest neighbors, forming a matrix $Z_i$ of size $n\times k$. The nearest neighbor search is done using a Euclidean metric, $d\left(\mathbf{z}_i, \mathbf{z}_j\right) = \left\|\mathbf{z}_i - \mathbf{z}_j\right\|_2^2$, limited to a search window of size $b \times b$ around the center of the patch. 

The matrix $Z_i$ undergoes a series of trainable operations that aim to recover a clean candidate patch $\mathbf{\hat{z}}_i$. Our filtering network consists of several blocks, each consisting of (i) a forward 2D linear transform; (ii) a non-negative thresholding (ReLU) on the obtained features; and (iii) a transform back to the image domain. All transforms operate separately in the spatial and the similarity domains. In contrast to BM3D and other methods, our filtering predicts the residual rather than the clean https://www.overleaf.com/project/5d49ef57a1ec5606cb231fc1patches, just as done in \cite{Zhang_DnCnn_2017}. This means that we estimate the noise using the network. Restored patches are obtained by subtracting the estimated noise from the corrupted patches. 

\begin{SCfigure}
    \centering
	\includegraphics[width=0.24\textwidth]{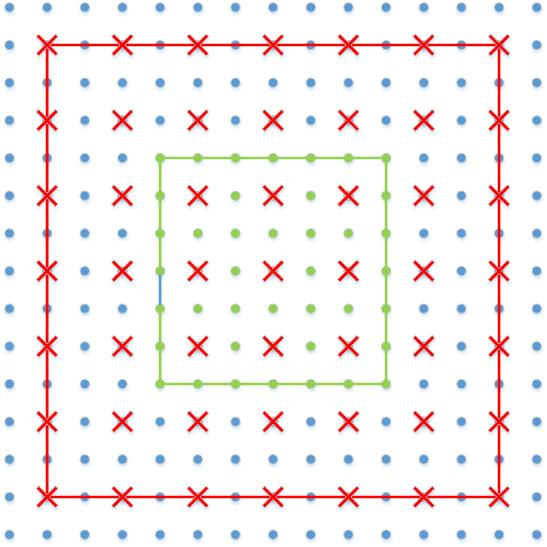}
	\caption{ Visualization of the corresponding 1$^\textrm{st}$ and 2$^\textrm{nd}$ scale patches: Blue dots are the processed image pixels; the green square is a 1$^\textrm{st}$ scale patch, while the red square is its corresponding 2$^\textrm{nd}$ scale patch. Both are of size $7 \times 7$ pixels.\vspace{16pt}}
	\label{fig:second_scale_patch}
\end{SCfigure}

Our scheme includes a multi-scale treatment and a fusion of corresponding patches from the different scales. The adopted strategy borrows its rationale from~\cite{Yang_SR_2010}, which studied the single image super resolution task. In their algorithm, high-resolution patches and their corresponding low-resolution versions are jointly treated by assuming that they share the same sparse representation. The two resolution patches are handled by learning a pair of coupled dictionaries. In a similar fashion, we augment the corresponding patches from the two scales, and learn a joint transform for their fusion. In our experiments, the multi-scale scheme includes only two scales, but the same concept can be applied to a higher pyramid. In our notations, the 1$^\textrm{st}$ scale is the original noisy image $\mathbf{Y}$, and the 2$^\textrm{nd}$ scale images are created by convolving $\mathbf{Y}$ with the low-pass filter
\begin{equation}
f_{LP} = \frac{1}{16}\begin{bmatrix} 1 & 2 & 1 \end{bmatrix}^T \cdot \begin{bmatrix} 1 & 2 & 1 \end{bmatrix}
\label{eq:f_lp}
\end{equation}
and down-sampling the result by a factor of two. In order to synchronize between patch locations in the two scales, we create four downscaled images by sampling the convolved image at either even or odd locations: $\mathbf{Y}_{ee}^{(2)}, \mathbf{Y}_{eo}^{(2)}, \mathbf{Y}_{oe}^{(2)}, \mathbf{Y}_{oo}^{(2)}$ (even/odd columns \& even/odd rows). For each 1$^\textrm{st}$ scale patch, the corresponding 2$^\textrm{nd}$ scale patch (of the same size $\sqrt{n} \times \sqrt{n}$) is extracted from the appropriate down-sampled image, such that both patches are centred at the same pixel in the original image, as depicted in Figure~\ref{fig:second_scale_patch}. 

We denote the 2$^\textrm{nd}$ scale patch that corresponds to  $\mathbf{z}_i$ by $\mathbf{z}_i^{(2)} \in \mathbb{R}^n$. This patch is augmented with a group of its $k - 1$ nearest neighbors, forming a matrix $Z_i^{(2)}$ of size $n \times k$. The nearest neighbor search is performed in the same down-scaled image from which $\mathbf{z}_i^{(2)}$ is taken, while limiting the search to a window of size $b \times b$. Both matrices, $\left\{Z_i\right\}$ and $\{Z_i^{(2)}\}$, are fed to the filtering network, which fuses the scales using a joint transform. The architecture of this network is described next.


\subsection{The Filtering Network}
\label{sec:architecture}

We turn to present the architecture of the filtering network, starting by describing the involved building blocks and then discussing the whole scheme. A basic component within our network is the TRT (Transform--ReLU--Transform) block. This follows the classic Thresholding algorithm in sparse approximation theory~\cite{Elad_Sparse_Book_2010} in which denoising is obtained by transforming the incoming signal, discarding of small entries (this way getting a sparse representation), and applying an inverse transform that returns to the signal domain. Note that the same conceptual structure is employed by the well-known BM3D algorithm~\cite{Egiazarian_BM3D_2007}. 
In a similar fashion, our TRT block applies a learned transform, non-negative thresholding (ReLU) and another transform on the resulting matrix. Both transforms are separable and linear, denoted by the operator $\operatorname{T}$ and implemented using a Separable Linear (SL) layer,
\begin{equation}
    \operatorname{T}\left(Z\right) = \operatorname{SL}\left(Z\right) = W_1 Z W_2 + B \;,
\end{equation}
where $W_1$ and $W_2$ operate in the spatial and the similarity domains respectively. Separability of the SL layer allows a substantial reduction in the number of parameters of the network. In fact, computing $W_1 Z W_2$ is equivalent to applying $W_{total}\mathbf{z}$, where $W_{total} = W_2^T \otimes W_1$ is a Kronecker product between $W_2^T$ and $W_1$, and $\mathbf{z}$ is a vectorized version of $Z$. 

\begin{figure}
    \centering
    \includegraphics[width=0.48\textwidth]{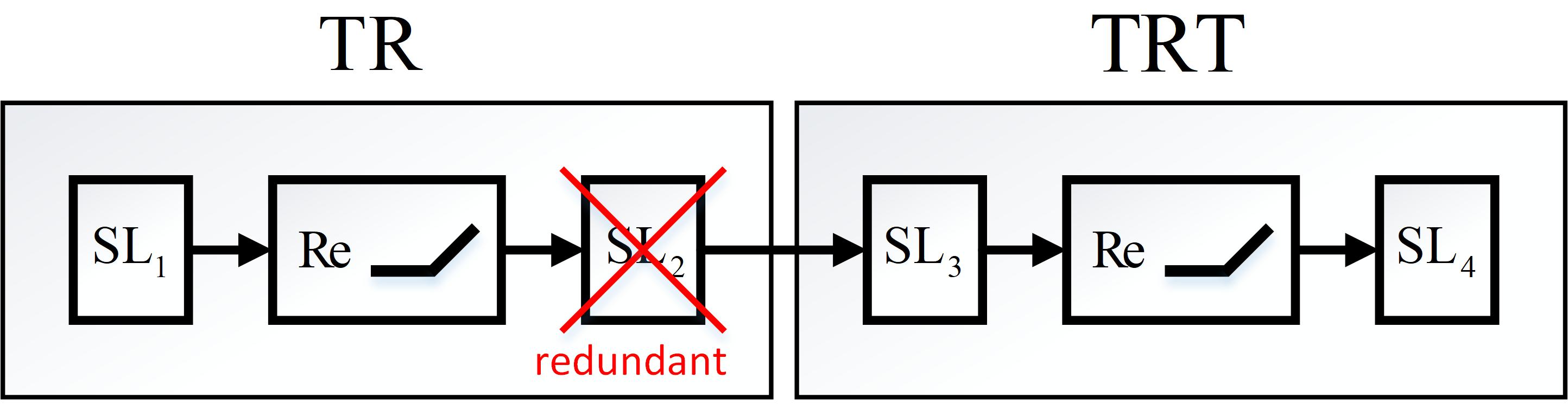}
    \caption{Concatenation of two TRTs: One SL layer is removed due to linearity, converting the first TRT block into TR.}
    \label{fig:tr_trt}
\end{figure}

Since concatenation of two $\operatorname{SL}$ layers can be replaced by a single effective $\operatorname{SL}$, due to their linearity, we remove one $\operatorname{SL}$ layer in any concatenation of two $\operatorname{TRT}$-s, as shown in Figure~\ref{fig:tr_trt}. 
The $\operatorname{TRT}$ component without the second transform is denoted by $\operatorname{TR}$, and when concatenating $k$ $\operatorname{TRT}$-s, the first $k - 1$ blocks should be replaced by $\operatorname{TR}$-s. Another variant we use in our network is $\operatorname{TBR}$, which is a version of $\operatorname{TR}$ with batch normalization added before the ReLU.

\begin{figure}
    \centering
    \includegraphics[width=0.48\textwidth]{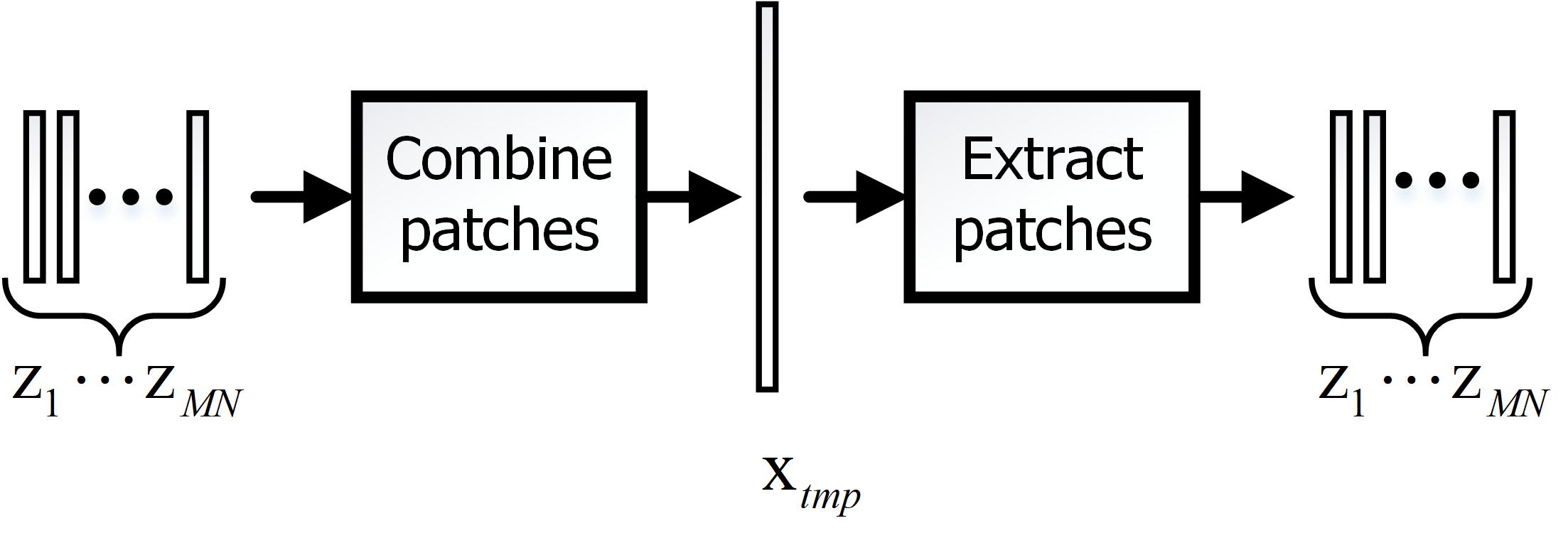}
    \caption{Aggregation ($\operatorname{AGG}$) block.}
    \label{fig:agg}
\end{figure}

Another component of the filtering network is an Aggregation block ($\operatorname{AGG}$), depicted in Figure~\ref{fig:agg}. 
This block imposes consistency between overlapping patches $\left\{\mathbf{z}_i\right\}$ by combining them to a temporary image $\mathbf{x}_{tmp}$ using plain averaging (as described in Eq. (\ref{eq:Agg}) but without the weights),  
and extracting them back from the obtained image by $R_i \mathbf{x}_{tmp}$.

\begin{figure*}
    \centering
    \includegraphics[width=0.96\textwidth]{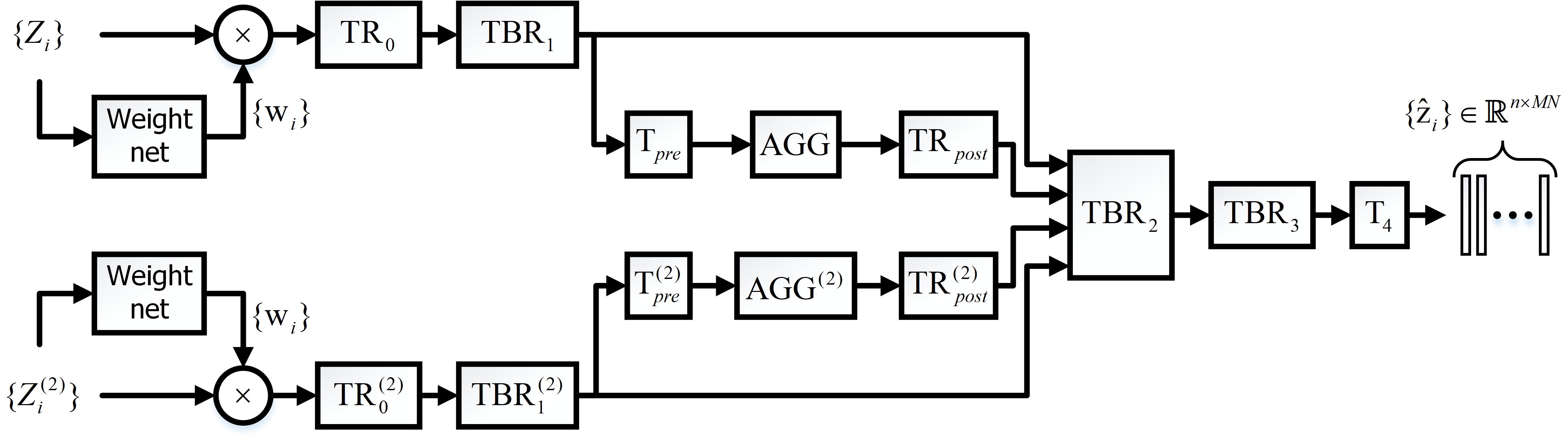}
    \renewcommand{\arraystretch}{1.4}
    \begin{tabular}{|c|c|c|c|c|c|c|c|} 
    \hline
    \multirow{2}{*}{} & $TR_0$ & $TBR_1$ & $T_{pre}$ & $TR_{post}$ & \multirow{2}{*}{$TBR_2$} & \multirow{2}{*}{$TBR_3$} & \multirow{2}{*}{$T_4$} \\
    \cline{2-5}
    & $TR_0^{(2)}$ & $TBR_1^{(2)}$ & $T_{pre}^{(2)}$ & $TR_{post}^{(2)}$ & & & \\
    \hline
    \renewcommand{\arraystretch}{1}
    $W_1$ & $49 \times 64$ & $64 \times 64$ & $64 \times 49$ & $49 \times 64$ & $64 \times 64$ & $64 \times 64$ & $64 \times 49$ \\ 
    \hline
    $W_2$ & $14 \times 14$ & $14 \times 14$ & $14 \times 1$ & $1 \times 14$ & $56 \times 56$ & $56 \times 56$ & $56 \times 1$ \\
    \hline
    \end{tabular}
    \renewcommand{\arraystretch}{1}
    \caption{The filtering network. The table below summarizes the sizes of the SL layers.}
    \label{fig:filtering_network}
\end{figure*}

The complete architecture of the filtering network is presented in Figure~\ref{fig:filtering_network}. The network receives as input two sets of matrices, $\{Z_i\}$ and $\{Z_i^{(2)}\}$, and its output is an array of filtered overlapping patches $\{\mathbf{\hat{z}}_i\}$. At first, each of these matrices is multiplied by a diagonal weight matrix $\operatorname{diag}(\mathbf{w}_i)Z_i$. 
Recall that the columns of $Z_i$ (or $\{Z_i^{(2)}\}$) are image patches, where the first is the processed patch and the rest are its neighbors. The weights $\mathbf{w}_i$ express the network's belief regarding the relevance of each of the neighbor patches to the denoising process. These weights are calculated using an auxiliary network denoted as ``weight net'', which consists of seven FC (Fully Connected) layers of size $k \times k$ with batch normalization and ReLU between each two FC's. The network gets as input the sample variance of the processed patch and $k - 1$ squared distances between the patch and its nearest neighbors. 

After multiplication by $\operatorname{diag}(\mathbf{w}_i)$ the $Z_i$ matrix undergoes a series of operations that include transforms, ReLUs and $\operatorname{AGG}$, until it gets to the $\operatorname{TBR}_2$ block, as shown in Figure~\ref{fig:filtering_network}. The aggregation block imposes consistency of the $\{Z_i\}$ matrices, which represent overlapping patches, but also causes loss of some inf.ormation, therefore we split the flow to two branches: with and without $\operatorname{AGG}$. Since the output of any $\operatorname{TR}$ or $\operatorname{TBR}$ component is in the feature domain, we wrap the $\operatorname{AGG}$ block with $\operatorname{T}_{pre}$ and $\operatorname{TR}_{post}$, where $\operatorname{T}_{pre}$ transforms the features to the image domain, and $\operatorname{TR}_{post}$ transforms the $\operatorname{AGG}$ output back to the feature space, while imposing sparsity. The 2$^\textrm{nd}$ scale matrices, $\{Z_i^{(2)}\}$, undergo very similar operations as the 1$^\textrm{st}$ scale ones, but with different learned parameters. The only difference in the treatment of the two scales is in the functionality of the aggregation blocks. Since the $\operatorname{AGG}^{(2)}$ operates on downsampled patches, combination and extraction is done with ${stride = 2}$. Additionally, $\operatorname{AGG}^{(2)}$ applies bilinear low-pass filter as defined in Equation~(\ref{eq:f_lp}) on the temporary image obtained after the patch combination.

The $\operatorname{TBR}_2$ block applies a joint transform that fuses the features coming from four origins: the 1$^\textrm{st}$ and 2$^\textrm{nd}$ scales with and without aggregation. The columns of all these matrices are concatenated together such that the same spatial transformation $W_1$ is applied on all. Note that the network size can be reduced at the cost of a slight degradation in performance by removing the  $\operatorname{TBR}_1$, $\operatorname{TBR}_1^{(2)}$ and $\operatorname{TBR}_3$ components. We discuss this option in the result section. 



\section{Image Adaptation}
\label{sec:adaptation}

The above designed network, once trained, offers a universal denoising machine, capable of removing white additive Gaussian noise from all images by applying the same set of computations to all images. As such, this machine is not adapted to the incoming image, which might deviate from the general statistics of the training corpus, or could have an inner structure that is not fully exploited. This lack of adaptation may imply reduced denoising performance. In this section we address this weakness by discussing an augmentation of our denoising algorithm that leads to such an adaptation and improved denoising results. 

Several recent papers have already proposed techniques for training networks while only using corrupted examples~\cite{Lehtinen_2018_Noise2noise,Ulyanov_DIP_2018,Krull_2019_Noise2void,Mataev_2019_Deepred,Batson_2019_Noise2self,Laine_2019_Self_Supervised_Denoising}. An interesting special case is where the network is trained on the corrupted image itself. Indeed, this single-image unsupervised learning has been successfully employed by classical algorithms. For example, The KSVD Denoising algorithm~\cite{Elad_Image_Denoising_KSVD_2006} trains a dictionary using patches extracted from the corrupted image. Another example is PLE~\cite{Sapiro_PLE_2012}, in which a GMM model is learned using the given image. Recent deep learning based methods~\cite{Ulyanov_DIP_2018,Mataev_2019_Deepred} adopted this approach, training on the corrupted image. However, their obtained performance tends to be non-competitive with fully supervised and universal schemes.

This raises an intriguing question: Could deep regression machines benefit from both supervised and single-image unsupervised training techniques? In this paper we provide a positive answer to this question, while leveraging the fact that our denoising network is lightweight. Our proposed denoiser is able to combine knowledge learned from an external dataset with knowledge that lies in the currently processed image, all while avoiding overfitting. We introduce two novel types of \emph{adaptation} techniques, external and internal, which should remind the reader of transfer learning. Both adaptation types start by denoising the input image regularly. Then the network is re-trained and updated for few epochs. In the external adaptation case, we seek (e.g., using Google image search) one or few other images closely related to the processed one, and re-train the network on this small set of clean images (and their noisy versions). In the internal adaptation mode of work, the network is re-trained on the denoised image itself using a loss function of the form
\begin{equation}
    \mathcal{L}\left(\theta\right) = \left\|\operatorname{f}_\theta \left(\hat{Y} + \mathbf{n}\right) - \hat{Y} \right\|^2_2 \;,
\end{equation}
where $\hat{Y} = \operatorname{f}_\theta \left(Y\right)$ is the universally denoised  image, and $\mathbf{n}$ stands for a synthetic additive Gaussian noise. For both external and internal adaptations the procedure concludes by denoising the input image by the updated network.

While the two adaptation methods seem similar, they serve different needs. In accordance with the above intuition, external adaptation should be chosen when handling images with special content that is not well represented in the training set. For example, this mode of work could be applied on non-natural images, such as scanned documents. Processing images with special content could be handled by training class-aware denoisers, however this approach might require a large amount of images for training each class, and holding many networks for covering the variety of classes to handle. For example, the work reported in~\cite{remez2017deep,remez2018class} train their networks on $900$ images per class. In contrast, our proposed external adaptation requires maintaining a single network for all images, while updating it for each incoming image. In Section~\ref{sec:adaptation_results} we present denoising experiments with non-natural images, in which applying external adaptation gains substantial improvement of more than $1$dB in terms of PSNR. In contrast to external adaptation, the internal one becomes effective when the incoming image is characterized by a high level of self-similarity. For example, as shown in Section~\ref{sec:adaptation_results}, applying internal adaptation on images from Urban~100~\cite{Huang_2015_Urban100}  gains a notable improvement of almost 0.3dB in PSNR on average. 

We should note that the proposed adaptation procedures are not always successful. However, failure cases usually do not cause performance degradation, indicating that these procedures, in the context of being deployed on a lightweight network, do not overfit. Indeed, while we got a negligible decrease in performance of up to $0.02$dB for few images with the internal adaptation, most unsuccessful tests led to a marginal increment of at least $0.02 - 0.05$dB.  


\section{Experimental Results: Universal Denoising}
\label{sec:results}

This section reports the  performance of the proposed scheme, with a comprehensive comparison to recent SOTA denoising algorithms. In particular, we include in these comparisons the classical  BM3D~\cite{Egiazarian_BM3D_2007} due to its resemblance to our network architecture, the TNRD~\cite{Chen_TNRD_2015}, DnCNN~\cite{Zhang_DnCnn_2017} and FFDNet~\cite{Zhang_FFDNet_2018} networks, the non-local and high performance NLRN~\cite{Liu_NLRN_2018} architecture, and the recently published Learned K-SVD (LKSVD)~\cite{Scetbon_LKSVD_2019} method. We also include comparisons to Lefkimiatis' networks, NLNet~\cite{Lefkimmiatis_NLNet_2017} and UNLNet~\cite{Lefkimmiatis_UNLNet_2018}, which  inspired our work. Our algorithm is denoted as Lightweight Learned Image Denoising with Instance Adaptation (LIDIA), and we present two versions of it, LIDIA and LIDIA-S. The second is a simplified network with slightly weaker performance (see more below).


\subsection{Denoising with Known Noise Level}
\label{sec:results:known_noise_level}

We start with plain denoising experiments, in which the noise is Gaussian white of known variance. This is the common case covered by all the above mentioned methods. 

Our network is trained on 432 images from the BSD500 set~\cite{martin2001_bsd_database}, and the evaluation uses the remaining 68 images (BSD68). The network is trained end-to-end using decreasing learning rate over batches of 4 images, using the mean-squared-error loss. We start training with the Adam optimizer with a learning rate of $10^{-2}$, and switch to SGD at the last part of the training with an initial learning rate of $10^{-3}$. 

\begin{figure}
    \centering
    \renewcommand{\arraystretch}{1.2}
    \includegraphics[width=0.48\textwidth]{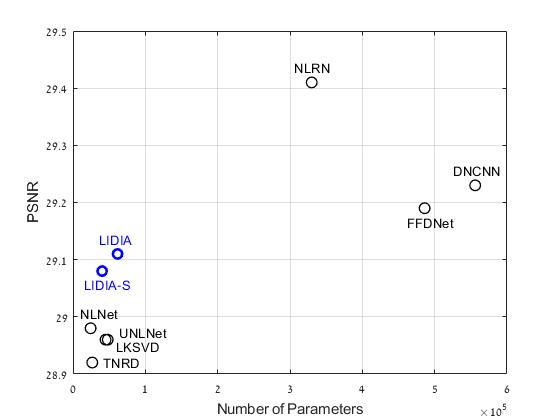}
    \caption{Comparing denoising networks: PSNR performance vs. the number of trained parameters (for noise level $\sigma = 25$).}
    \label{fig:psnr_vs_parameters}
\end{figure}

Figure~\ref{fig:psnr_vs_parameters} presents a comparison between our algorithm and leading alternative ones by presenting their PSNR performance versus their number of trained parameters. This figure exposes the fact that the performance of denoising networks is heavily influenced by their complexity. As can be seen, the various algorithms can be roughly split into two categories: lightweight architectures with a number of parameters below 100K (TNRD~\cite{Chen_TNRD_2015}, LKSVD~\cite{Scetbon_LKSVD_2019}, NLNet~\cite{Lefkimmiatis_NLNet_2017} and UNLNet\footnote{UNLNet is a blind denoising network trained for ${5 \le \sigma < 30}$.}~\cite{Lefkimmiatis_UNLNet_2018}), and much larger and slightly better performing networks (DnCNN~\cite{Zhang_DnCnn_2017}, FFDNet~\cite{Zhang_FFDNet_2018}, and NLRN~\cite{Liu_NLRN_2018}) that use hundreds of thousands of parameters. As we proceed in this section, we emphasize lightweight architectures in our comparisons, a category to which our network belongs. Figure~\ref{fig:psnr_vs_parameters} shows that our networks (both LIDIA and LIDIA-S) achieve the best results within this lightweight category. 

\begin{figure*}
    \centering
	\begin{subfigure}{0.24\textwidth}
	    \captionsetup{justification=centering}
		\includegraphics[width=\textwidth]{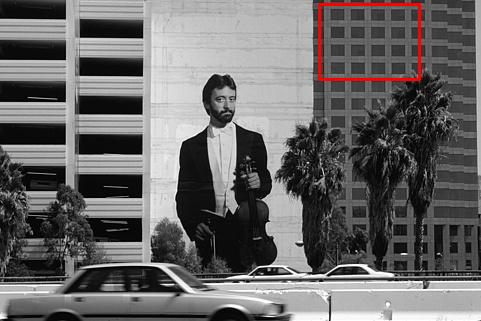}
		\caption{Original \newline}
		\label{fig:clean011}
		\vspace*{6pt}
	\end{subfigure}
	\begin{subfigure}{0.24\textwidth}
	    \captionsetup{justification=centering}
		\includegraphics[width=\textwidth]{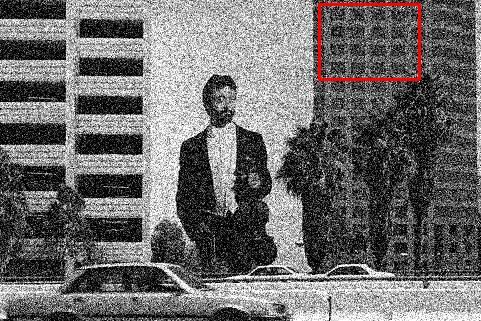}
		\caption{Noisy with $\sigma = 50$ \newline}
		\label{fig:noisy011_sig50}
		\vspace*{6pt}
	\end{subfigure}
	\begin{subfigure}{0.24\textwidth}
	    \captionsetup{justification=centering}
		\includegraphics[width=\textwidth]{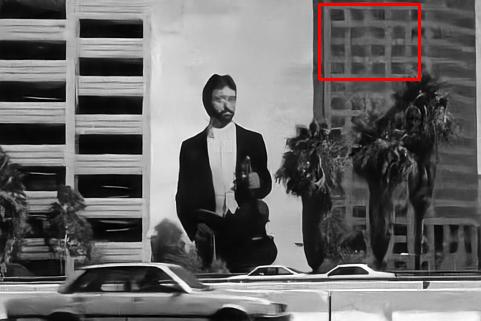}
		\caption{DnCNN~\cite{Zhang_DnCnn_2017} \\ PSNR = 25.56dB}
		\label{fig:dncnn011_sig50}
		\vspace*{6pt}
	\end{subfigure}
	\begin{subfigure}{0.24\textwidth}
	    \captionsetup{justification=centering}
		\includegraphics[width=\textwidth]{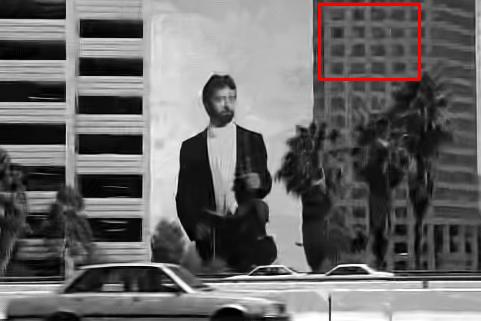}
		\caption{BM3D~\cite{Egiazarian_BM3D_2007} \\ PSNR = 24.99dB}
		\label{fig:bm3d011_sig50}
		\vspace*{6pt}
	\end{subfigure} \\
	\begin{subfigure}{0.24\textwidth}
	    \captionsetup{justification=centering}
		\includegraphics[width=\textwidth]{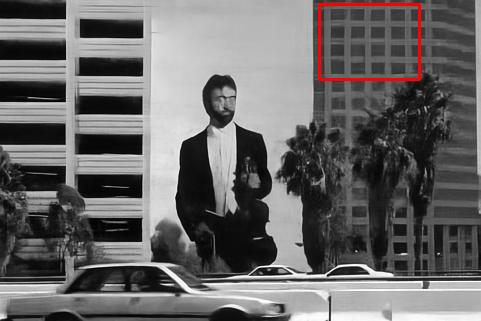}
		\caption{NLRN~\cite{Liu_NLRN_2018} \\ PSNR = 26.11dB}
		\label{fig:nlrn011_sig50}
		\vspace*{6pt}
	\end{subfigure}
	\begin{subfigure}{0.24\textwidth}
	    \captionsetup{justification=centering}
		\includegraphics[width=\textwidth]{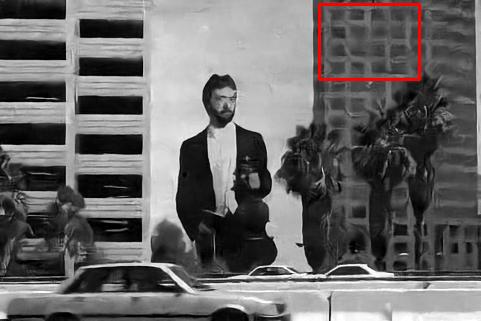}
		\caption{TNRD~\cite{Chen_TNRD_2015} \\ PSNR = 25.07dB}
		\label{fig:tnrd011_sig50}
		\vspace*{6pt}
	\end{subfigure}
	\begin{subfigure}{0.24\textwidth}
	    \captionsetup{justification=centering}
		\includegraphics[width=\textwidth]{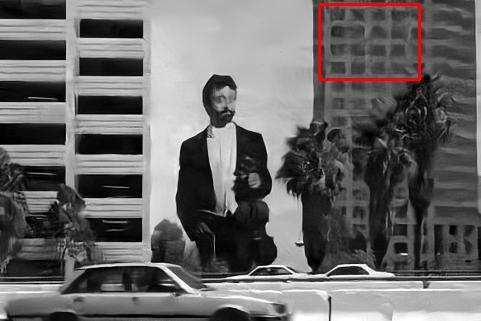}
		\caption{NLNet~\cite{Lefkimmiatis_NLNet_2017} \\ PSNR = 25.21dB}
		\label{fig:nlnet011_sig50}
		\vspace*{6pt}
	\end{subfigure}
	\begin{subfigure}{0.24\textwidth}
	    \captionsetup{justification=centering}
		\includegraphics[width=\textwidth]{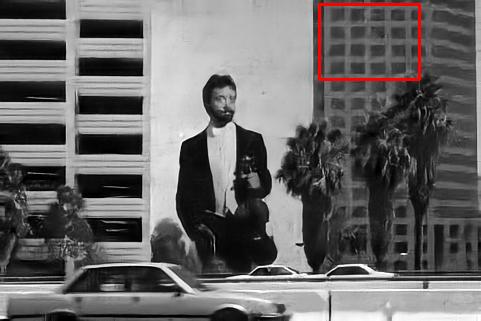}
		\caption{LIDIA (ours) \\ PSNR = 25.60dB}
		\label{fig:nlms011_sig50}
		\vspace*{6pt}
	\end{subfigure} \\
	\begin{subfigure}{0.24\textwidth}
	    \captionsetup{justification=centering}
		\includegraphics[width=\textwidth]{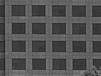}
		\caption{Original}
		\label{fig:clean011_zoom}
		\vspace*{6pt}
	\end{subfigure}
	\begin{subfigure}{0.24\textwidth}
	    \captionsetup{justification=centering}
		\includegraphics[width=\textwidth]{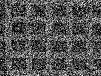}
		\caption{Noisy}
		\label{fig:noisy011_zoom_sig50}
		\vspace*{6pt}
	\end{subfigure}
	\begin{subfigure}{0.24\textwidth}
	    \captionsetup{justification=centering}
		\includegraphics[width=\textwidth]{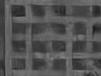}
		\caption{DnCNN}
		\label{fig:dncnn011_zoom_sig50}
		\vspace*{6pt}
	\end{subfigure}
	\begin{subfigure}{0.24\textwidth}
	    \captionsetup{justification=centering}
		\includegraphics[width=\textwidth]{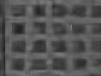}
		\caption{BM3D}
		\label{fig:bm3d011_zoom_sig50}
		\vspace*{6pt}
	\end{subfigure} \\
	\begin{subfigure}{0.24\textwidth}
	    \captionsetup{justification=centering}
		\includegraphics[width=\textwidth]{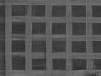}
		\caption{NLRN}
		\label{fig:nlrn011_zoom_sig50}
		\vspace*{6pt}
	\end{subfigure}
	\begin{subfigure}{0.24\textwidth}
	    \captionsetup{justification=centering}
		\includegraphics[width=\textwidth]{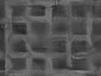}
		\caption{TNRD}
		\label{fig:tnrd011_zoom_sig50}
		\vspace*{6pt}
	\end{subfigure}
	\begin{subfigure}{0.24\textwidth}
	    \captionsetup{justification=centering}
		\includegraphics[width=\textwidth]{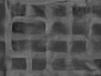}
		\caption{NLNet}
		\label{fig:nlnet011_zoom_sig50}
		\vspace*{6pt}
	\end{subfigure}
	\begin{subfigure}{0.24\textwidth}
	    \captionsetup{justification=centering}
		\includegraphics[width=\textwidth]{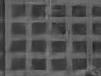}
		\caption{LIDIA (ours)}
		\label{fig:nlms011_zoom_sig50}
		\vspace*{6pt}
	\end{subfigure}
	\caption{Denoising example with $\sigma = 50$.}
	\label{fig:test011_sig50}
\end{figure*}

\begin{figure*}
    \centering
	\begin{subfigure}{0.24\textwidth}
	    \captionsetup{justification=centering}
		\includegraphics[width=\textwidth]{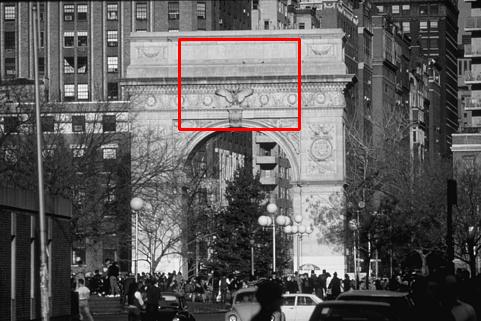}
		\caption{Original \newline}
		\label{fig:clean022}
		\vspace*{6pt}
	\end{subfigure}
	\begin{subfigure}{0.24\textwidth}
	    \captionsetup{justification=centering}
		\includegraphics[width=\textwidth]{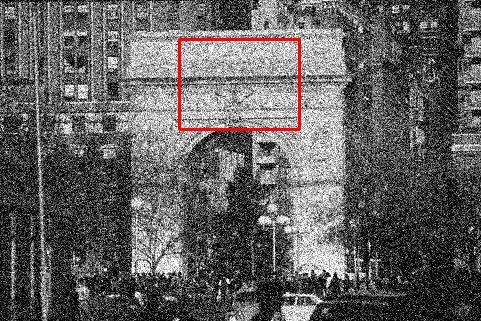}
		\caption{Noisy with $\sigma = 50$ \newline}
		\label{fig:noisy022_sig50}
		\vspace*{6pt}
	\end{subfigure}
	\begin{subfigure}{0.24\textwidth}
	    \captionsetup{justification=centering}
		\includegraphics[width=\textwidth]{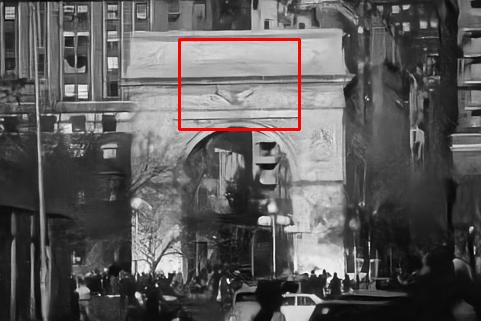}
		\caption{DnCNN~\cite{Zhang_DnCnn_2017} \\ PSNR = 23.95dB}
		\label{fig:dncnn022_sig50}
		\vspace*{6pt}
	\end{subfigure}
	\begin{subfigure}{0.24\textwidth}
	    \captionsetup{justification=centering}
		\includegraphics[width=\textwidth]{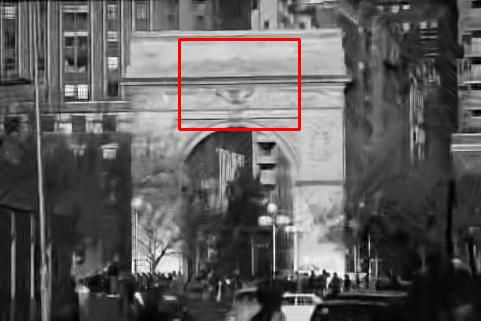}
		\caption{BM3D~\cite{Egiazarian_BM3D_2007} \\ PSNR = 23.36dB}
		\label{fig:bm3d022_sig50}
		\vspace*{6pt}
	\end{subfigure} \\
	\begin{subfigure}{0.24\textwidth}
	    \captionsetup{justification=centering}
		\includegraphics[width=\textwidth]{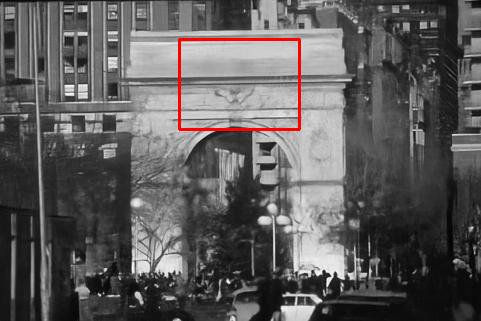}
		\caption{NLRN~\cite{Liu_NLRN_2018} \\ PSNR = 24.22dB}
		\label{fig:nlrn022_sig50}
		\vspace*{6pt}
	\end{subfigure}
	\begin{subfigure}{0.24\textwidth}
	    \captionsetup{justification=centering}
		\includegraphics[width=\textwidth]{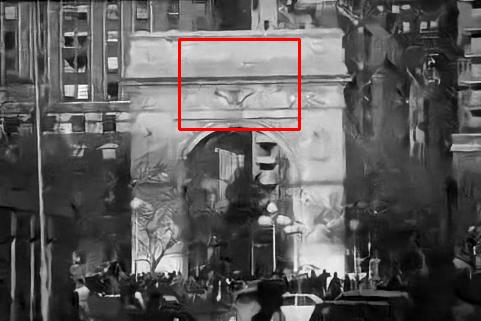}
		\caption{TNRD~\cite{Chen_TNRD_2015} \\ PSNR = 23.61dB}
		\label{fig:tnrd022_sig50}
		\vspace*{6pt}
	\end{subfigure}
	\begin{subfigure}{0.24\textwidth}
	    \captionsetup{justification=centering}
		\includegraphics[width=\textwidth]{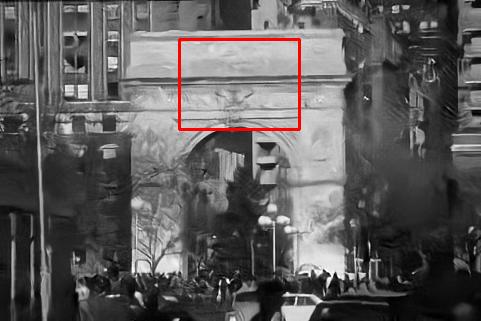}
		\caption{NLNet~\cite{Lefkimmiatis_NLNet_2017} \\ PSNR = 23.63dB}
		\label{fig:nlnet022_sig50}
		\vspace*{6pt}
	\end{subfigure}
	\begin{subfigure}{0.24\textwidth}
	    \captionsetup{justification=centering}
		\includegraphics[width=\textwidth]{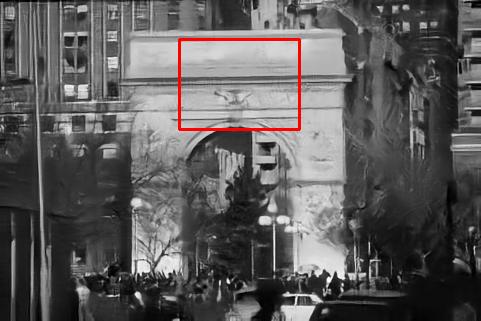}
		\caption{LIDIA (ours) \\ PSNR = 23.91dB}
		\label{fig:nlms022_sig50}
		\vspace*{6pt}
	\end{subfigure} \\
	\begin{subfigure}{0.24\textwidth}
	    \captionsetup{justification=centering}
		\includegraphics[width=\textwidth]{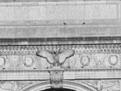}
		\caption{Original}
		\label{fig:clean022_zoom}
		\vspace*{6pt}
	\end{subfigure}
	\begin{subfigure}{0.24\textwidth}
	    \captionsetup{justification=centering}
		\includegraphics[width=\textwidth]{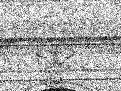}
		\caption{Noisy}
		\label{fig:noisy022_zoom_sig50}
		\vspace*{6pt}
	\end{subfigure}
	\begin{subfigure}{0.24\textwidth}
	    \captionsetup{justification=centering}
		\includegraphics[width=\textwidth]{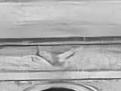}
		\caption{DnCNN}
		\label{fig:dncnn022_zoom_sig50}
		\vspace*{6pt}
	\end{subfigure}
	\begin{subfigure}{0.24\textwidth}
	    \captionsetup{justification=centering}
		\includegraphics[width=\textwidth]{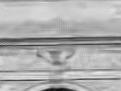}
		\caption{BM3D}
		\label{fig:bm3d022_zoom_sig50}
		\vspace*{6pt}
	\end{subfigure} \\
	\begin{subfigure}{0.24\textwidth}
	    \captionsetup{justification=centering}
		\includegraphics[width=\textwidth]{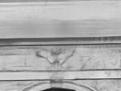}
		\caption{NLRN}
		\label{fig:nlrn022_zoom_sig50}
		\vspace*{6pt}
	\end{subfigure}
	\begin{subfigure}{0.24\textwidth}
	    \captionsetup{justification=centering}
		\includegraphics[width=\textwidth]{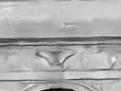}
		\caption{TNRD}
		\label{fig:tnrd022_zoom_sig50}
		\vspace*{6pt}
	\end{subfigure}
	\begin{subfigure}{0.24\textwidth}
	    \captionsetup{justification=centering}
		\includegraphics[width=\textwidth]{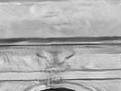}
		\caption{NLNet}
		\label{fig:nlnet022_zoom_sig50}
		\vspace*{6pt}
	\end{subfigure}
	\begin{subfigure}{0.24\textwidth}
	    \captionsetup{justification=centering}
		\includegraphics[width=\textwidth]{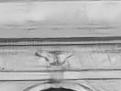}
		\caption{LIDIA (ours)}
		\label{fig:nlms022_zoom_sig50}
		\vspace*{6pt}
	\end{subfigure}
	\caption{Denoising example with $\sigma = 50$.}
	\label{fig:test022_sig50}
\end{figure*}

\begin{figure*}
    \centering
	\begin{subfigure}{0.24\textwidth}
	    \captionsetup{justification=centering}
		\includegraphics[width=\textwidth]{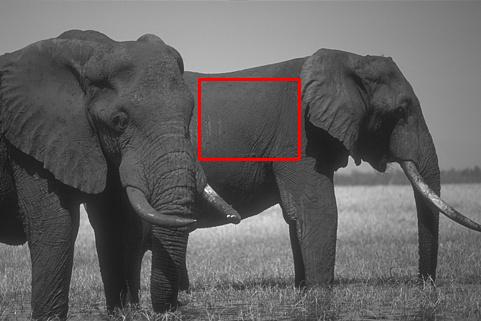}
		\caption{Original \newline}
		\label{fig:clean059}
		\vspace*{6pt}
	\end{subfigure}
	\begin{subfigure}{0.24\textwidth}
	    \captionsetup{justification=centering}
		\includegraphics[width=\textwidth]{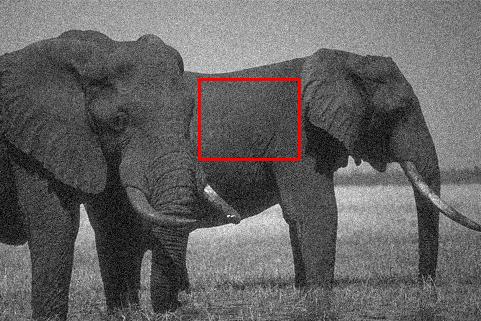}
		\caption{Noisy with $\sigma = 15$ \newline}
		\label{fig:noisy059_sig15}
		\vspace*{6pt}
	\end{subfigure}
	\begin{subfigure}{0.24\textwidth}
	    \captionsetup{justification=centering}
		\includegraphics[width=\textwidth]{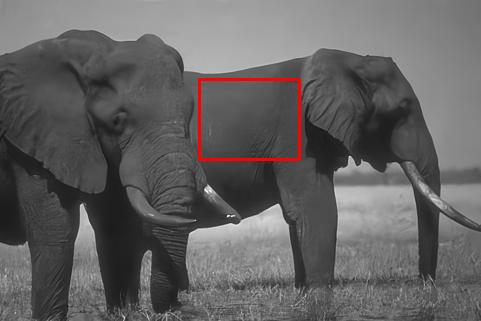}
		\caption{DnCNN~\cite{Zhang_DnCnn_2017} \\ PSNR = 32.32dB}
		\label{fig:dncnn059_sig15}
		\vspace*{6pt}
	\end{subfigure}
	\begin{subfigure}{0.24\textwidth}
	    \captionsetup{justification=centering}
		\includegraphics[width=\textwidth]{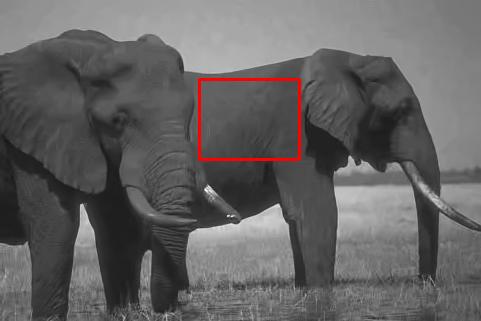}
		\caption{BM3D~\cite{Egiazarian_BM3D_2007} \\ PSNR = 31.70dB}
		\label{fig:bm3d059_sig15}
		\vspace*{6pt}
	\end{subfigure} \\
	\begin{subfigure}{0.24\textwidth}
	    \captionsetup{justification=centering}
		\includegraphics[width=\textwidth]{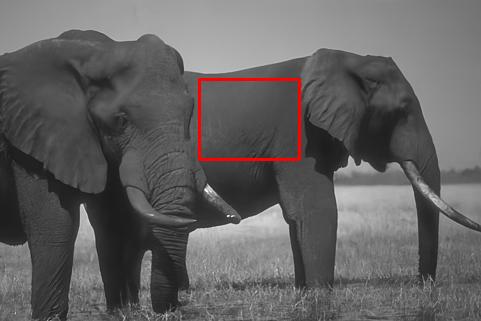}
		\caption{NLRN~\cite{Liu_NLRN_2018} \\ PSNR = 32.47dB}
		\label{fig:nlrn059_sig15}
		\vspace*{6pt}
	\end{subfigure}
	\begin{subfigure}{0.24\textwidth}
	    \captionsetup{justification=centering}
		\includegraphics[width=\textwidth]{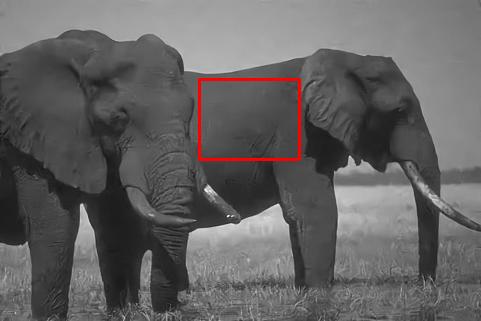}
		\caption{TNRD~\cite{Chen_TNRD_2015} \\ PSNR = 32.05dB}
		\label{fig:tnrd059_sig15}
		\vspace*{6pt}
	\end{subfigure}
	\begin{subfigure}{0.24\textwidth}
	    \captionsetup{justification=centering}
		\includegraphics[width=\textwidth]{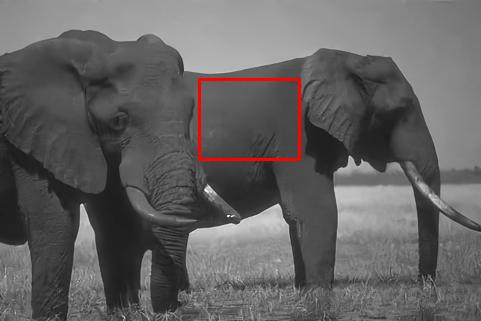}
		\caption{NLNet~\cite{Lefkimmiatis_NLNet_2017} \\ PSNR = 32.14dB}
		\label{fig:nlnet059_sig15}
		\vspace*{6pt}
	\end{subfigure}
	\begin{subfigure}{0.24\textwidth}
	    \captionsetup{justification=centering}
		\includegraphics[width=\textwidth]{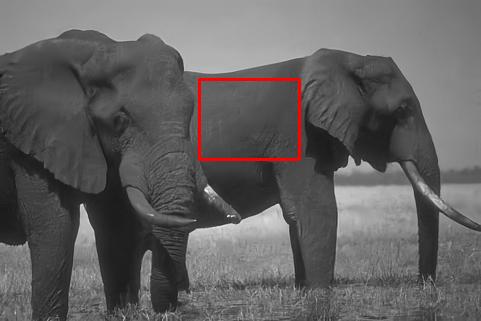}
		\caption{LIDIA (ours) \\ PSNR = 32.30dB}
		\label{fig:nlms059_sig15}
		\vspace*{6pt}
	\end{subfigure} \\
	\begin{subfigure}{0.24\textwidth}
	    \captionsetup{justification=centering}
		\includegraphics[width=\textwidth]{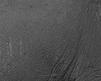}
		\caption{Original}
		\label{fig:clean059_zoom}
		\vspace*{6pt}
	\end{subfigure}
	\begin{subfigure}{0.24\textwidth}
	    \captionsetup{justification=centering}
		\includegraphics[width=\textwidth]{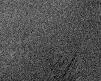}
		\caption{Noisy}
		\label{fig:noisy059_zoom_sig15}
		\vspace*{6pt}
	\end{subfigure}
	\begin{subfigure}{0.24\textwidth}
	    \captionsetup{justification=centering}
		\includegraphics[width=\textwidth]{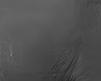}
		\caption{DnCNN}
		\label{fig:dncnn059_zoom_sig15}
		\vspace*{6pt}
	\end{subfigure}
	\begin{subfigure}{0.24\textwidth}
	    \captionsetup{justification=centering}
		\includegraphics[width=\textwidth]{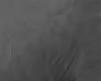}
		\caption{BM3D}
		\label{fig:bm3d059_zoom_sig15}
		\vspace*{6pt}
	\end{subfigure} \\
	\begin{subfigure}{0.24\textwidth}
	    \captionsetup{justification=centering}
		\includegraphics[width=\textwidth]{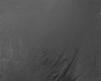}
		\caption{NLRN}
		\label{fig:nlrn059_zoom_sig15}
		\vspace*{6pt}
	\end{subfigure}
	\begin{subfigure}{0.24\textwidth}
	    \captionsetup{justification=centering}
		\includegraphics[width=\textwidth]{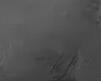}
		\caption{TNRD}
		\label{fig:tnrd059_zoom_sig15}
		\vspace*{6pt}
	\end{subfigure}
	\begin{subfigure}{0.24\textwidth}
	    \captionsetup{justification=centering}
		\includegraphics[width=\textwidth]{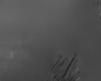}
		\caption{NLNet}
		\label{fig:nlnet059_zoom_sig15}
		\vspace*{6pt}
	\end{subfigure}
	\begin{subfigure}{0.24\textwidth}
	    \captionsetup{justification=centering}
		\includegraphics[width=\textwidth]{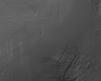}
		\caption{LIDIA (ours)}
		\label{fig:nlms059_zoom_sig15}
		\vspace*{6pt}
	\end{subfigure}
	\caption{Denoising example with $\sigma = 15$.}
	\label{fig:test059_sig15}
\end{figure*}

\begin{figure*}
    \centering
	\begin{subfigure}{0.21\textwidth}
	    \captionsetup{justification=centering}
		\includegraphics[width=\textwidth]{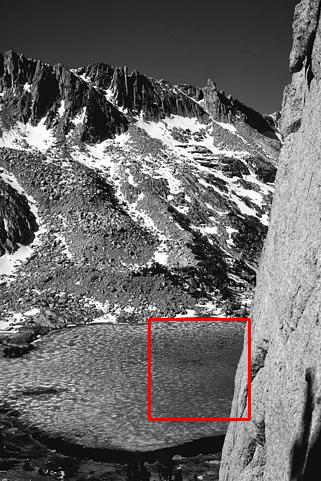}
		\caption{Original \newline}
		\label{fig:clean029}
		\vspace*{6pt}
	\end{subfigure}
	\begin{subfigure}{0.21\textwidth}
	    \captionsetup{justification=centering}
		\includegraphics[width=\textwidth]{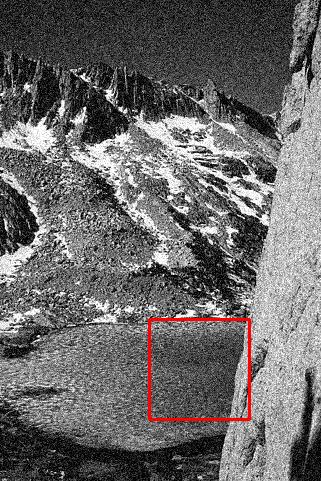}
		\caption{Noisy with $\sigma = 25$ \newline}
		\label{fig:noisy029_sig25}
		\vspace*{6pt}
	\end{subfigure}
	\begin{subfigure}{0.21\textwidth}
	    \captionsetup{justification=centering}
		\includegraphics[width=\textwidth]{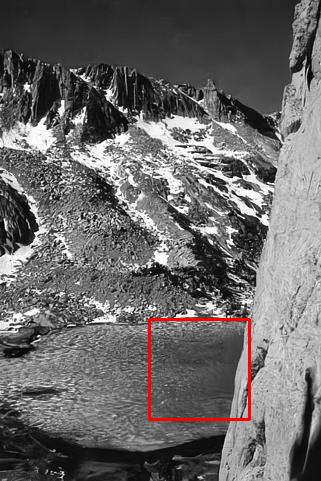}
		\caption{DnCNN~\cite{Zhang_DnCnn_2017} \\ PSNR = 24.47dB}
		\label{fig:dncnn029_sig25}
		\vspace*{6pt}
	\end{subfigure}
	\begin{subfigure}{0.21\textwidth}
	    \captionsetup{justification=centering}
		\includegraphics[width=\textwidth]{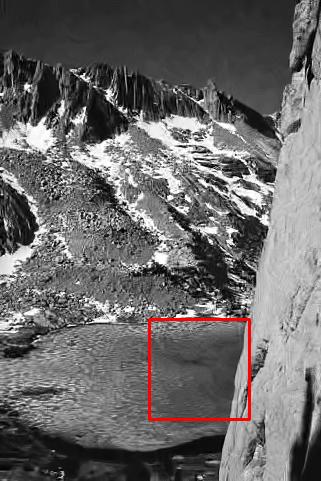}
		\caption{BM3D~\cite{Egiazarian_BM3D_2007} \\ PSNR = 23.81dB}
		\label{fig:bm3d029_sig25}
		\vspace*{6pt}
	\end{subfigure} \\
	\begin{subfigure}{0.21\textwidth}
	    \captionsetup{justification=centering}
		\includegraphics[width=\textwidth]{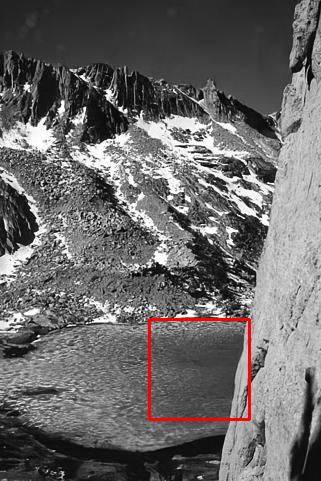}
		\caption{NLRN~\cite{Liu_NLRN_2018} \\ PSNR = 24.58dB}
		\label{fig:nlrn029_sig25}
		\vspace*{6pt}
	\end{subfigure}
	\begin{subfigure}{0.21\textwidth}
	    \captionsetup{justification=centering}
		\includegraphics[width=\textwidth]{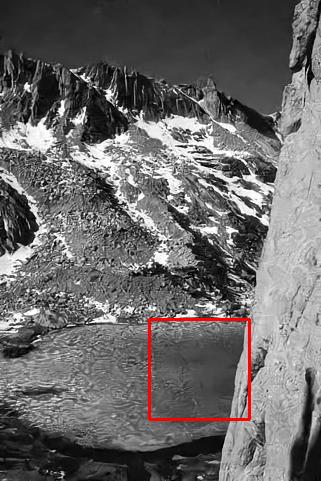}
		\caption{TNRD~\cite{Chen_TNRD_2015} \\ PSNR = 24.14dB}
		\label{fig:tnrd029_sig25}
		\vspace*{6pt}
	\end{subfigure}
	\begin{subfigure}{0.21\textwidth}
	    \captionsetup{justification=centering}
		\includegraphics[width=\textwidth]{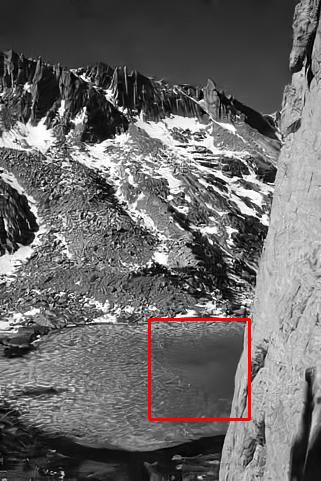}
		\caption{NLNet~\cite{Lefkimmiatis_NLNet_2017} \\ PSNR = 24.12dB}
		\label{fig:nlnet029_sig25}
		\vspace*{6pt}
	\end{subfigure}
	\begin{subfigure}{0.21\textwidth}
	    \captionsetup{justification=centering}
		\includegraphics[width=\textwidth]{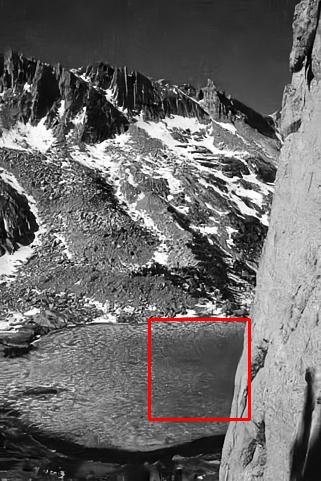}
		\caption{LIDIA (ours) \\ PSNR = 24.38dB}
		\label{fig:nlms029_sig25}
		\vspace*{6pt}
	\end{subfigure} \\
	\begin{subfigure}{0.21\textwidth}
	    \captionsetup{justification=centering}
		\includegraphics[width=\textwidth]{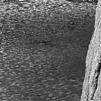}
		\caption{Original}
		\label{fig:clean029_zoom}
		\vspace*{6pt}
	\end{subfigure}
	\begin{subfigure}{0.21\textwidth}
	    \captionsetup{justification=centering}
		\includegraphics[width=\textwidth]{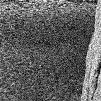}
		\caption{Noisy}
		\label{fig:noisy029_zoom_sig25}
		\vspace*{6pt}
	\end{subfigure}
	\begin{subfigure}{0.21\textwidth}
	    \captionsetup{justification=centering}
		\includegraphics[width=\textwidth]{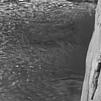}
		\caption{DnCNN}
		\label{fig:dncnn029_zoom_sig25}
		\vspace*{6pt}
	\end{subfigure}
	\begin{subfigure}{0.21\textwidth}
	    \captionsetup{justification=centering}
		\includegraphics[width=\textwidth]{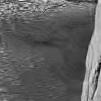}
		\caption{BM3D}
		\label{fig:bm3d029_zoom_sig25}
		\vspace*{6pt}
	\end{subfigure} \\
	\begin{subfigure}{0.21\textwidth}
	    \captionsetup{justification=centering}
		\includegraphics[width=\textwidth]{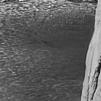}
		\caption{NLRN}
		\label{fig:nlrn029_zoom_sig25}
		\vspace*{6pt}
	\end{subfigure}
	\begin{subfigure}{0.21\textwidth}
	    \captionsetup{justification=centering}
		\includegraphics[width=\textwidth]{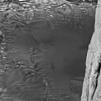}
		\caption{TNRD}
		\label{fig:tnrd029_zoom_sig25}
		\vspace*{6pt}
	\end{subfigure}
	\begin{subfigure}{0.21\textwidth}
	    \captionsetup{justification=centering}
		\includegraphics[width=\textwidth]{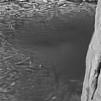}
		\caption{NLNet}
		\label{fig:nlnet029_zoom_sig25}
		\vspace*{6pt}
	\end{subfigure}
	\begin{subfigure}{0.21\textwidth}
	    \captionsetup{justification=centering}
		\includegraphics[width=\textwidth]{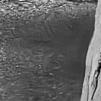}
		\caption{LIDIA (ours)}
		\label{fig:nlms029_zoom_sig25}
		\vspace*{6pt}
	\end{subfigure}
	\caption{Denoising example with $\sigma = 25$.}
	\label{fig:test029_sig25}
\end{figure*}

Detailed quantitative denoising results per noise level are reported\footnote{ NLNet~\cite{Lefkimmiatis_NLNet_2017} complexity and PSNR are taken from the released code.} in Table~\ref{tab:denoising_known_sig}. For each noise level, the best denoising performance is marked in red, and the best performance within the lightweight category is marked in blue. Table~\ref{tab:num_params} reports the number of trained parameters per each of the competing networks. Figures~\ref{fig:test011_sig50}, \ref{fig:test022_sig50}, \ref{fig:test059_sig15} and \ref{fig:test029_sig25} present examples of denoising results. Since our architecture is related to both BM3D and NLNet, we focus on qualitative comparisons with these algorithms. For all noise levels our results are significantly sharper, contain less artifacts and preserve more details than those of BM3D. In comparison to NLNet, our method is significantly better in high noise levels due to our multi-scale treatment, recovering large and repeating elements, as shown in Figure~\ref{fig:nlms011_zoom_sig50}. In fact our algorithm manages to recover repeating elements better than all methods presented in Figure~\ref{fig:test011_sig50} except NLRN. 
In addition, in cases of high noise levels, the multi-scale treatment allows handling smooth areas with less artifacts than NLNet, as one can see from the results in Figure \ref{fig:nlms022_zoom_sig50} and \ref{fig:nlnet022_zoom_sig50}. In medium noise levels, our algorithm recovers more details, while NLNet tends to over-smooth the recovered image. For example, see the Elephant skin in Figure~\ref{fig:test059_sig15} and the mountain glacier in Figure~\ref{fig:test029_sig25}. 

\begin{table}
    \centering
    \renewcommand{\arraystretch}{1.2}
    \setlength{\doublerulesep}{1pt}
    \begin{tabular}{|c||c|c|c||c|}
        \hline
        \multirow{2}{*}{Method} & \multicolumn{3}{c||}{Noise $\sigma$} & \multirow{2}{*}{Average} \\
        \hhline{|~||---||~}
        & 15 & 25 & 50 & \\
        \hhline{|=#=|=|=#=|}
        $\operatorname{TNRD}$\cite{Chen_TNRD_2015} & 31.42 & 28.92 & 25.97 & 28.77 \\
        \hline
        $\operatorname{DnCNN}$\cite{Zhang_DnCnn_2017} & 31.73 & 29.23 & 26.23 & 29.06 \\
        \hline
        $\operatorname{BM3D}$\cite{Egiazarian_BM3D_2007} & 31.07 & 28.57 & 25.62 & 28.42 \\
        \hline
        $\operatorname{NLRN}$\cite{Liu_NLRN_2018} & \textcolor{red}{31.88} & \textcolor{red}{29.41} & \textcolor{red}{26.47} & \textcolor{red}{29.25} \\
        \hline
        $\operatorname{NLNet}$\footnotemark[1]\cite{Lefkimmiatis_NLNet_2017} & 31.50 & 28.98 & 26.03 & 28.84 \\
        \hline
        $\operatorname{LIDIA}$ (ours) & \textcolor{blue}{31.62} & \textcolor{blue}{29.11} & \textcolor{blue}{26.17} & \textcolor{blue}{28.97} \\
        \hline
    \end{tabular}
    \caption{B/W Denoising performance with known noise level: Best PSNR is marked in \textcolor{red}{red}, and best PSNR within the lightweight category is marked in \textcolor{blue}{blue}.}
    \label{tab:denoising_known_sig}
\end{table}


\begin{table}
    \centering
    \begin{tabular}{|c|c|c|c|c|c|c|}
        \hline
        $\operatorname{DnCNN}$ & $\operatorname{NLRN}$ & $\operatorname{TNRD}$ & $\operatorname{NLNet}$\footnotemark[1] & $\operatorname{LIDIA}$ & $\operatorname{LIDIA-S}$ \\
         \hline
        556K & 330K & 26.6K & 24.3K & 61.6K & 40.2K \\
         \hline
    \end{tabular}
    \caption{Denoising networks: Number of parameters.}
    \label{tab:num_params}
\end{table}

\begin{figure*}
    \centering
	\begin{subfigure}{0.24\textwidth}
	    \captionsetup{justification=centering}
		\includegraphics[width=\textwidth]{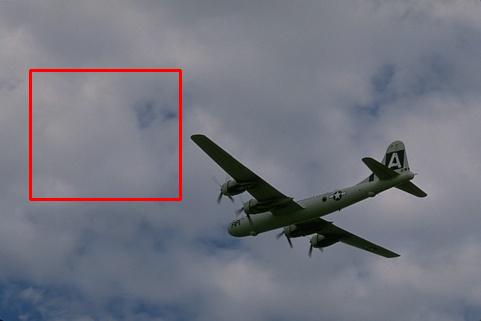}
		\caption{Original \newline}
		\label{fig:clean3096}
		\vspace*{6pt}
	\end{subfigure}
	\begin{subfigure}{0.24\textwidth}
	    \captionsetup{justification=centering}
		\includegraphics[width=\textwidth]{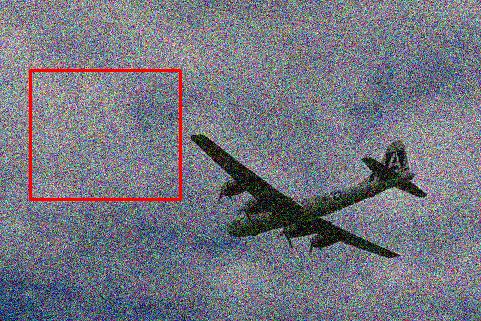}
		\caption{Noisy with $\sigma = 50$ \newline}
		\label{fig:noisy3096_sig50}
		\vspace*{6pt}
	\end{subfigure}
	\begin{subfigure}{0.24\textwidth}
	    \captionsetup{justification=centering}
		\includegraphics[width=\textwidth]{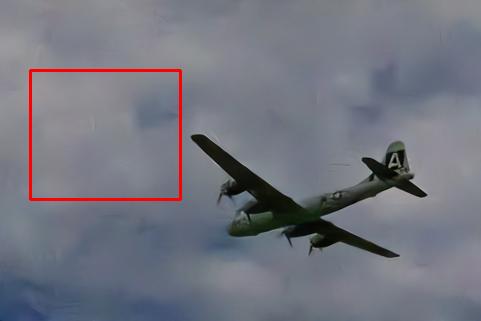}
		\caption{CDnCNN~\cite{Zhang_DnCnn_2017} \\ PSNR = 36.68dB}
		\label{fig:dncnn3096_sig50}
		\vspace*{6pt}
	\end{subfigure}
	\begin{subfigure}{0.24\textwidth}
	    \captionsetup{justification=centering}
		\includegraphics[width=\textwidth]{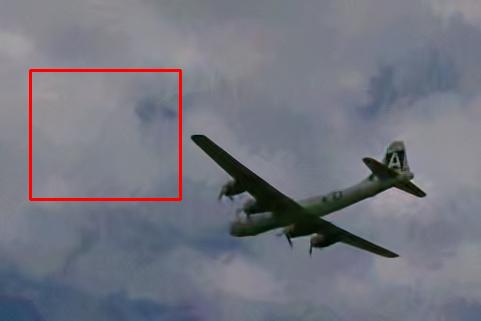}
		\caption{CBM3D~\cite{dabov2007color} \\ PSNR = 35.51dB}
		\label{fig:bm3d3096_sig50}
		\vspace*{6pt}
	\end{subfigure} \\
	\begin{subfigure}{0.24\textwidth}
	    \captionsetup{justification=centering}
		\includegraphics[width=\textwidth]{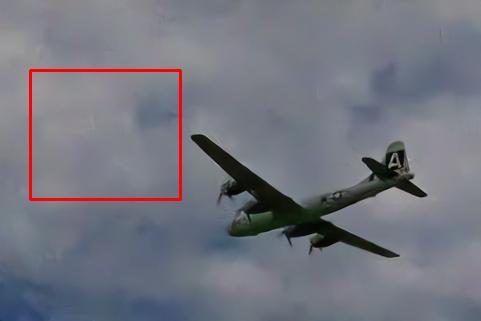}
		\caption{CFFDNet~\cite{Zhang_FFDNet_2018} \\ PSNR = 36.62dB}
		\label{fig:ffdnet3096_sig50}
		\vspace*{6pt}
	\end{subfigure}
	\begin{subfigure}{0.24\textwidth}
	    \captionsetup{justification=centering}
		\includegraphics[width=\textwidth]{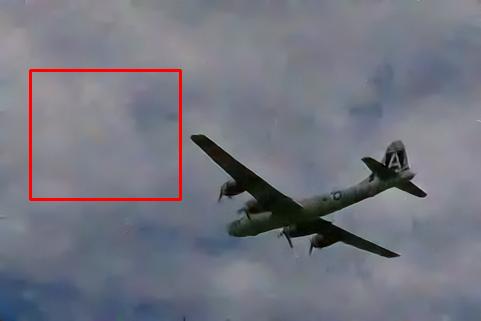}
		\caption{CNLNet~\cite{Lefkimmiatis_NLNet_2017} \\ PSNR = 35.98dB}
		\label{fig:nlnet3096_sig50}
		\vspace*{6pt}
	\end{subfigure}
	\begin{subfigure}{0.24\textwidth}
	    \captionsetup{justification=centering}
		\includegraphics[width=\textwidth]{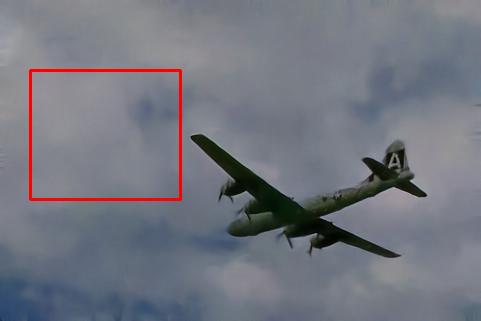}
		\caption{C-LIDIA (ours) \\ PSNR = 36.56dB}
		\label{fig:nlms3096_sig50}
		\vspace*{6pt}
	\end{subfigure} \\
	\begin{subfigure}{0.24\textwidth}
	    \captionsetup{justification=centering}
		\includegraphics[width=\textwidth]{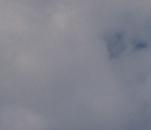}
		\caption{Original}
		\label{fig:clean3096_zoom}
		\vspace*{6pt}
	\end{subfigure}
	\begin{subfigure}{0.24\textwidth}
	    \captionsetup{justification=centering}
		\includegraphics[width=\textwidth]{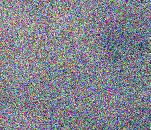}
		\caption{Noisy}
		\label{fig:noisy3096_zoom_sig50}
		\vspace*{6pt}
	\end{subfigure}
	\begin{subfigure}{0.24\textwidth}
	    \captionsetup{justification=centering}
		\includegraphics[width=\textwidth]{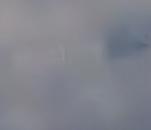}
		\caption{CDnCNN}
		\label{fig:dncnn3096_zoom_sig50}
		\vspace*{6pt}
	\end{subfigure}
	\begin{subfigure}{0.24\textwidth}
	    \captionsetup{justification=centering}
		\includegraphics[width=\textwidth]{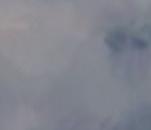}
		\caption{CBM3D}
		\label{fig:bm3d3096_zoom_sig50}
		\vspace*{6pt}
	\end{subfigure} \\
	\begin{subfigure}{0.24\textwidth}
	    \captionsetup{justification=centering}
		\includegraphics[width=\textwidth]{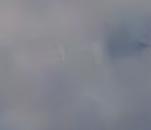}
		\caption{CFFDNet}
		\label{fig:ffdnet3096_zoom_sig50}
		\vspace*{6pt}
	\end{subfigure}
	\begin{subfigure}{0.24\textwidth}
	    \captionsetup{justification=centering}
		\includegraphics[width=\textwidth]{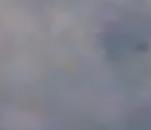}
		\caption{CNLNet}
		\label{fig:nlnet3096_zoom_sig50}
		\vspace*{6pt}
	\end{subfigure}
	\begin{subfigure}{0.24\textwidth}
	    \captionsetup{justification=centering}
		\includegraphics[width=\textwidth]{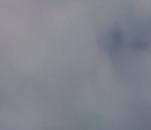}
		\caption{C-LIDIA (ours)}
		\label{fig:nlms3096_zoom_sig50}
		\vspace*{6pt}
	\end{subfigure}
	\caption{Color Image denoising example with $\sigma = 50$.}
	\label{fig:test3096_sig50}
\end{figure*}

\begin{figure*}
    \centering
	\begin{subfigure}{0.24\textwidth}
	    \captionsetup{justification=centering}
		\includegraphics[width=\textwidth]{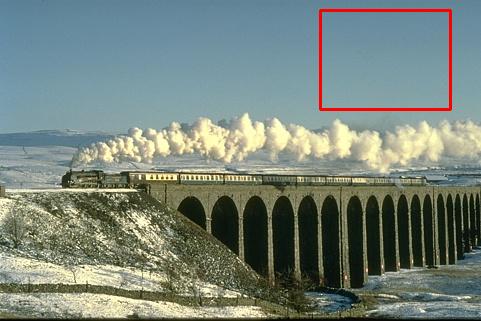}
		\caption{Original \newline}
		\label{fig:clean182053}
		\vspace*{6pt}
	\end{subfigure}
	\begin{subfigure}{0.24\textwidth}
	    \captionsetup{justification=centering}
		\includegraphics[width=\textwidth]{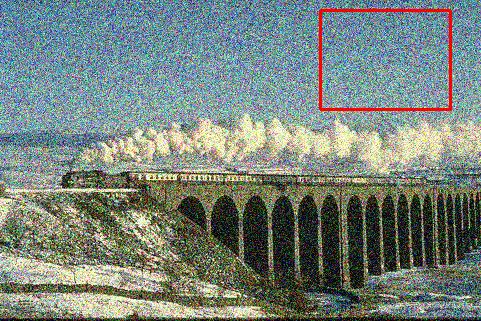}
		\caption{Noisy with $\sigma = 50$ \newline}
		\label{fig:noisy182053_sig50}
		\vspace*{6pt}
	\end{subfigure}
	\begin{subfigure}{0.24\textwidth}
	    \captionsetup{justification=centering}
		\includegraphics[width=\textwidth]{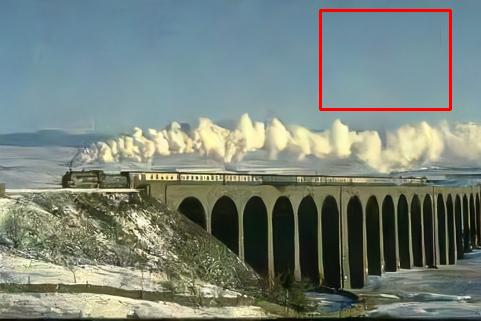}
		\caption{CDnCNN~\cite{Zhang_DnCnn_2017} \\ PSNR = 26.85dB}
		\label{fig:dncnn182053_sig50}
		\vspace*{6pt}
	\end{subfigure}
	\begin{subfigure}{0.24\textwidth}
	    \captionsetup{justification=centering}
		\includegraphics[width=\textwidth]{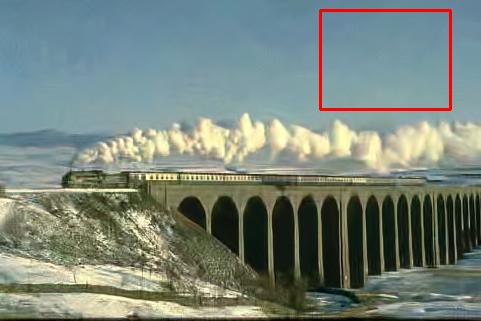}
		\caption{CBM3D~\cite{dabov2007color} \\ PSNR = 26.23dB}
		\label{fig:bm3d182053_sig50}
		\vspace*{6pt}
	\end{subfigure} \\
	\begin{subfigure}{0.24\textwidth}
	    \captionsetup{justification=centering}
		\includegraphics[width=\textwidth]{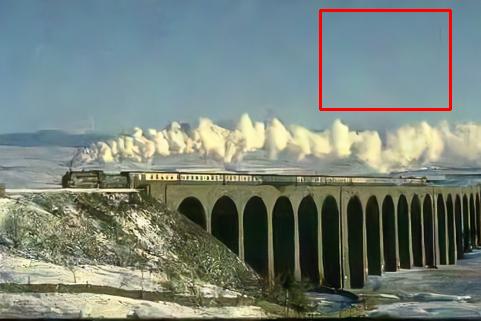}
		\caption{CFFDNet~\cite{Zhang_FFDNet_2018} \\ PSNR = 26.78dB}
		\label{fig:ffdnet182053_sig50}
		\vspace*{6pt}
	\end{subfigure}
	\begin{subfigure}{0.24\textwidth}
	    \captionsetup{justification=centering}
		\includegraphics[width=\textwidth]{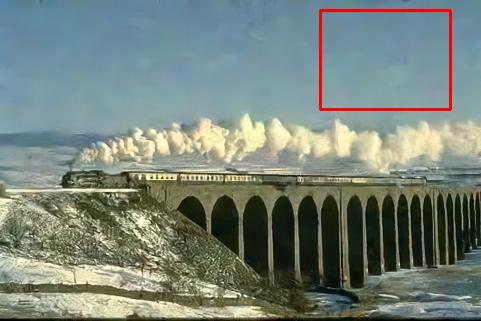}
		\caption{CNLNet~\cite{Lefkimmiatis_NLNet_2017} \\ PSNR = 26.62dB}
		\label{fig:nlnet182053_sig50}
		\vspace*{6pt}
	\end{subfigure}
	\begin{subfigure}{0.24\textwidth}
	    \captionsetup{justification=centering}
		\includegraphics[width=\textwidth]{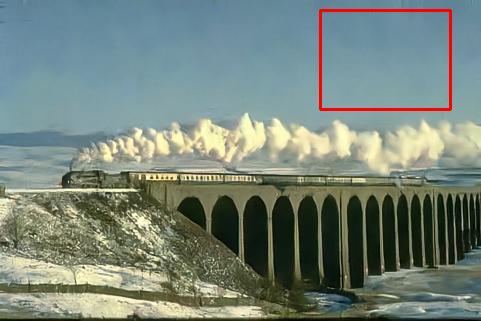}
		\caption{C-LIDIA (ours) \\ PSNR = 26.90dB}
		\label{fig:nlms182053_sig50}
		\vspace*{6pt}
	\end{subfigure} \\
	\begin{subfigure}{0.24\textwidth}
	    \captionsetup{justification=centering}
		\includegraphics[width=\textwidth]{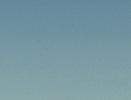}
		\caption{Original}
		\label{fig:clean182053_zoom}
		\vspace*{6pt}
	\end{subfigure}
	\begin{subfigure}{0.24\textwidth}
	    \captionsetup{justification=centering}
		\includegraphics[width=\textwidth]{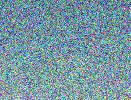}
		\caption{Noisy}
		\label{fig:noisy182053_zoom_sig50}
		\vspace*{6pt}
	\end{subfigure}
	\begin{subfigure}{0.24\textwidth}
	    \captionsetup{justification=centering}
		\includegraphics[width=\textwidth]{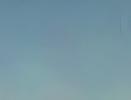}
		\caption{CDnCNN}
		\label{fig:dncnn182053_zoom_sig50}
		\vspace*{6pt}
	\end{subfigure}
	\begin{subfigure}{0.24\textwidth}
	    \captionsetup{justification=centering}
		\includegraphics[width=\textwidth]{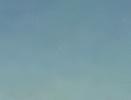}
		\caption{CBM3D}
		\label{fig:bm3d182053_zoom_sig50}
		\vspace*{6pt}
	\end{subfigure} \\
	\begin{subfigure}{0.24\textwidth}
	    \captionsetup{justification=centering}
		\includegraphics[width=\textwidth]{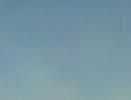}
		\caption{CFFDNet}
		\label{fig:ffdnet182053_zoom_sig50}
		\vspace*{6pt}
	\end{subfigure}
	\begin{subfigure}{0.24\textwidth}
	    \captionsetup{justification=centering}
		\includegraphics[width=\textwidth]{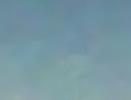}
		\caption{CNLNet}
		\label{fig:nlnet182053_zoom_sig50}
		\vspace*{6pt}
	\end{subfigure}
	\begin{subfigure}{0.24\textwidth}
	    \captionsetup{justification=centering}
		\includegraphics[width=\textwidth]{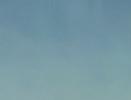}
		\caption{C-LIDIA (ours)}
		\label{fig:nlms182053_zoom_sig50}
		\vspace*{6pt}
	\end{subfigure}
	\caption{Color image denoising example with $\sigma = 50$.}
	\label{fig:test182053_sig50}
\end{figure*}

For denoising of color images we use 3D patches of size $5 \times 5 \times 3$ and increase the size of the matrices $W_1$ from ($49 \times 64$, $64 \times 64$, $64 \times 49$) to ($75 \times 80$, $80 \times 80$, $80 \times 75$) accordingly, which increases the total number of our network's parameters to 94K. The nearest neighbor search is done using the Luminance component, $Y = 0.2989 R + 0.5870 G + 0.1140 B $. Quantitative denoising results are reported in Table~\ref{tab:color_denoising}, where our network is denoted as C-LIDIA (Color LIDIA). As can be seen, our network is the best within the lightweight category, and gets quite close to the CDnCNN performance~\cite{Zhang_DnCnn_2017}. Figures~\ref{fig:test3096_sig50}~and~\ref{fig:test182053_sig50} present examples of denoising results which show that C-LIDIA handles low frequency noise better than CBM3D and CNLNet due to its multi-scale treatment.

\begin{table}
    \centering
    \renewcommand{\arraystretch}{1.2}
    \setlength{\doublerulesep}{1pt}
    \begin{tabular}{|c||c|c|c||c|}
        \hline
        \multirow{2}{*}{Method} & \multicolumn{3}{c||}{Noise $\sigma$} & \multirow{2}{*}{Average} \\
        \hhline{|~||---||~}
        & 15 & 25 & 50 & \\
        \hhline{|=#=|=|=#=|}
        $\operatorname{CFFDNet}$\cite{Zhang_FFDNet_2018} & 33.87 & 31.21 & 27.96 & 31.01 \\
        \hline
        $\operatorname{CDnCNN}$\cite{Zhang_DnCnn_2017} & 33.99 & \textcolor{red}{31.31} & \textcolor{red}{28.01} & 31.10 \\
        \hline
        $\operatorname{CBM3D}$\cite{dabov2007color} & 33.50 & 30.68 & 27.36 & 30.51 \\
        \hline
        $\operatorname{CNLNet}$\footnotemark[1]\cite{Lefkimmiatis_NLNet_2017} & 33.81 & 31.08 & 27.73 & 30.87 \\
        \hline
        $\operatorname{C-LIDIA}$(our) & \textcolor{red}{34.03} & \textcolor{red}{31.31} & \textcolor{blue}{27.99} & \textcolor{red}{31.11} \\
        \hline
    \end{tabular}
    \caption{Color image denoising performance: Best PSNR is marked in \textcolor{red}{red}, and best PSNR within the lightweight category is marked in \textcolor{blue}{blue}.}
    \label{tab:color_denoising}
\end{table}

\subsection{Blind Denoising}

Blind denoising, i.e., denoising with unknown noise level, is a useful feature when it comes to neural networks. This allows using a fixed network for performing image denoising, while serving a range of noise levels. This is a more practical solution, when compared to the one discussed above, in which we have designed a series of networks, each trained for a particular $\sigma$. We report blind denoising performance of our architecture and compare to similar results by \hbox{DnCNN-b}~\cite{Zhang_DnCnn_2017} (a version of DnCNN that has been trained for a range of $\sigma$ values) and UNLNet~\cite{Lefkimmiatis_UNLNet_2018}. Our blind denoising network (denoted LIDIA-b) preserves all its structure, but simply trained by mixing noise level examples in the range $10 \le \sigma \le 30$. The evaluation of all three networks is performed on images with $\sigma = [15, 25]$. The results of this experiment are  brought in Table~\ref{tab:blind_denoising}. As can be seen, our method obtains a higher PSNR than UNLNet, while being slightly weaker than DnCNN-b. Considering again the fact that our network has nearly $10\%$ of the parameters of DnCNN-b, we can say that our approach leads to SOTA results in the lightweight category.

\begin{table}
    \centering
    \renewcommand{\arraystretch}{1.2}
    \setlength{\doublerulesep}{1pt}
    \begin{tabular}{|c||c|c||c|}
        \hline
        \multirow{2}{*}{Method} & \multicolumn{2}{c||}{Noise $\sigma$} & \multirow{2}{*}{Average} \\
        \hhline{|~||--||~}
        & 15 & 25 & \\
        \hhline{|=#=|=#=|}
        $\operatorname{DnCNN-b}$\cite{Zhang_DnCnn_2017} & 31.61 & 29.16 & 30.39 \\
        \hline
        $\operatorname{UNLNet}$\cite{Lefkimmiatis_UNLNet_2018} & 31.47 & 28.96 & 30.22 \\
        \hline
        $\operatorname{LIDIA-b}$ (ours) & 31.54 & 29.06 & 30.30 \\
        \hline
    \end{tabular}
    \caption{Blind denoising performance.}
    \label{tab:blind_denoising}
\end{table}


\subsection{Reducing Network Size}

\begin{figure*}
    \centering
	\begin{subfigure}{0.24\textwidth}
	    \captionsetup{justification=centering}
		\includegraphics[width=\textwidth]{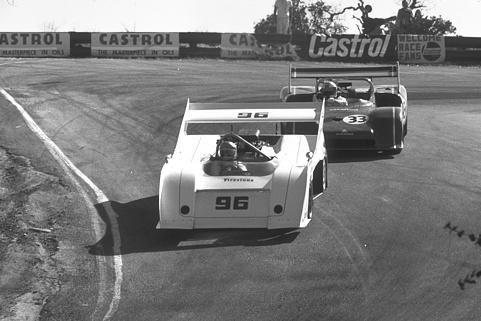}
		\caption{Original \newline \newline}
		\label{fig:clean040}
		\vspace*{6pt}
	\end{subfigure}
	\begin{subfigure}{0.24\textwidth}
	    \captionsetup{justification=centering}
		\includegraphics[width=\textwidth]{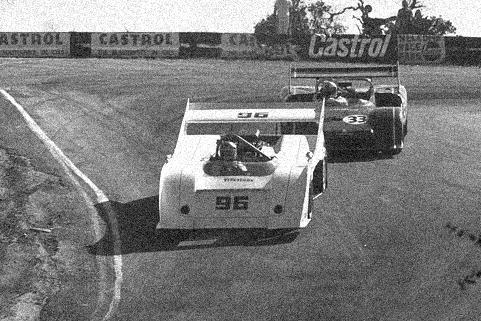}
		\caption{Noisy with $\sigma = 15$ \newline \newline}
		\label{fig:noisy040_sig15}
		\vspace*{6pt}
	\end{subfigure}
	\begin{subfigure}{0.24\textwidth}
	    \captionsetup{justification=centering}
		\includegraphics[width=\textwidth]{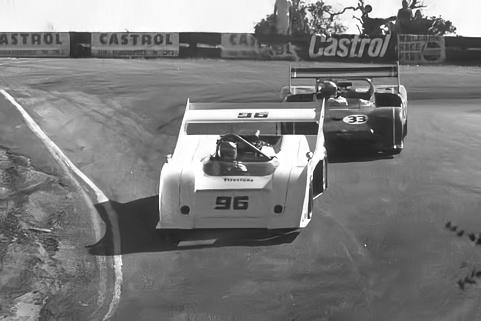}
		\caption{LIDIA \\ PSNR = 29.38dB \newline}
		\label{fig:nlms040_sig15}
		\vspace*{6pt}
	\end{subfigure}
	\begin{subfigure}{0.24\textwidth}
	    \captionsetup{justification=centering}
		\includegraphics[width=\textwidth]{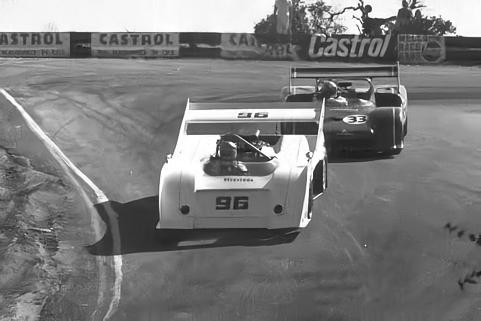}
		\caption{LIDIA-S \\ PSNR = 29.34dB \newline}
		\label{fig:nlms_s040_sig15}
		\vspace*{6pt}
	\end{subfigure} \\
	\begin{subfigure}{0.24\textwidth}
	    \captionsetup{justification=centering}
		\includegraphics[width=\textwidth]{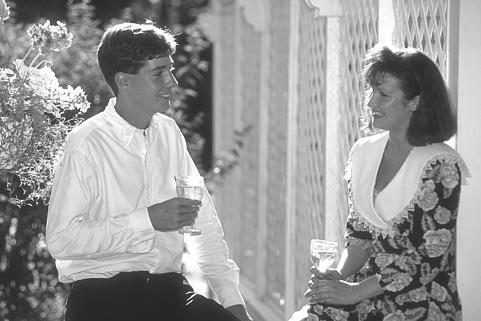}
		\caption{Original \newline \newline}
		\label{fig:clean023}
		\vspace*{6pt}
	\end{subfigure}
	\begin{subfigure}{0.24\textwidth}
	    \captionsetup{justification=centering}
		\includegraphics[width=\textwidth]{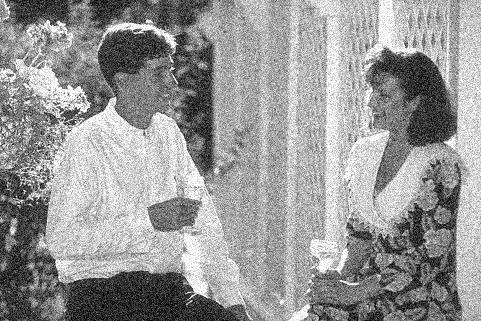}
		\caption{Noisy with $\sigma = 25$ \newline \newline}
		\label{fig:noisy023_sig25}
		\vspace*{6pt}
	\end{subfigure}
	\begin{subfigure}{0.24\textwidth}
	    \captionsetup{justification=centering}
		\includegraphics[width=\textwidth]{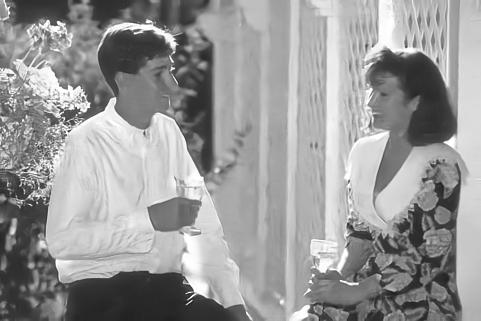}
		\caption{LIDIA \\ PSNR = 28.96dB \newline}
		\label{fig:nlms023_sig25}
		\vspace*{6pt}
	\end{subfigure}
	\begin{subfigure}{0.24\textwidth}
	    \captionsetup{justification=centering}
		\includegraphics[width=\textwidth]{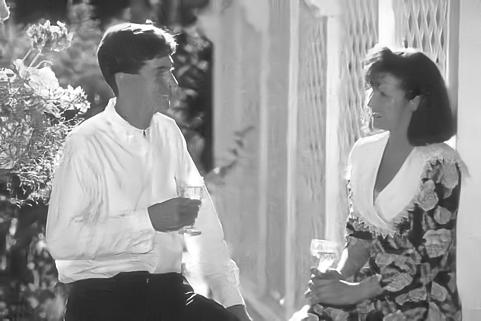}
		\caption{LIDIA-S \\ PSNR = 28.89dB \newline}
		\label{fig:nlms_s023_sig25}
		\vspace*{6pt}
	\end{subfigure}
	\caption{Comparison between full and small versions of the LIDIA network.}
	\label{fig:nlms_vs_nlms_s}
\end{figure*}

Our LIDIA denoising network can be further simplified by removing the $\operatorname{TBR}_1$, $\operatorname{TBR}_1^{(2)}$ and $\operatorname{TBR}_3$ components. The resulting smaller network, denoted by LIDIA-S, contains 30\% less parameters than the original LIDIA architecture (see Table~\ref{tab:num_params}), while achieving slightly weaker performance. Table~\ref{tab:full_vs_small} shows that for both regular and blind denoising scenarios, LIDIA-S  achieves an average PSNR that is only $0.05$dB lower than the full-size LIDIA network. Denoising examples are presented in Figure~\ref{fig:nlms_vs_nlms_s}, showing that the visual quality gap between LIDIA and LIDIA-S is marginal.

\begin{table}
    \centering
    \renewcommand{\arraystretch}{1.2}
    \setlength{\doublerulesep}{1pt}
    \begin{tabular}{|c||c|c|c||c|}
        \hline
        \multirow{2}{*}{Method} & \multicolumn{3}{c||}{Noise $\sigma$} & \multirow{2}{*}{Average} \\
        \cline{2-4}
        & 15 & 25 & 50 & \\
        \hline
        $\operatorname{LIDIA}$ & 31.62 & 29.11 & 26.17 & 28.97 \\
        \hline
        $\operatorname{LIDIA-S}$ & 31.57 & 29.08 & 26.13 & 28.93 \\
        \hline
        $\operatorname{LIDIA-b}$ & 31.54 & 29.06 & -- & 30.30 \\
        \hline
        $\operatorname{LIDIA-S-b}$ & 31.49 & 29.01 & -- & ‭30.25 ‬\\
        \hline
    \end{tabular}
    \caption{Performance comparison between LIDIA and its smaller version, LIDIA-S.}
    \label{tab:full_vs_small}
\end{table}

\section{Experimental Results: Network Adaptation}
\label{sec:adaptation_results}

We now turn to present results related to external and internal image adaptation. In our experiments we compare the performance of our network (before and after adaptation) with DnCNN~\cite{Zhang_DnCnn_2017}, one of the best deeply-learned denoisers. Unless said otherwise, all adaptation results are obtained via $5$ training epochs, requiring several minutes (depending on the image size) on Nvidia GeForce GTX 1080 Ti GPU. The adaptation does not require early stopping, i.e., training the network over tens of epochs leads to similar, and sometimes even better results.


\subsection{External Adaptation}

\begin{figure}
    \centering
	\begin{subfigure}{0.2\textwidth}
	    \captionsetup{justification=centering}
		\includegraphics[width=\textwidth]{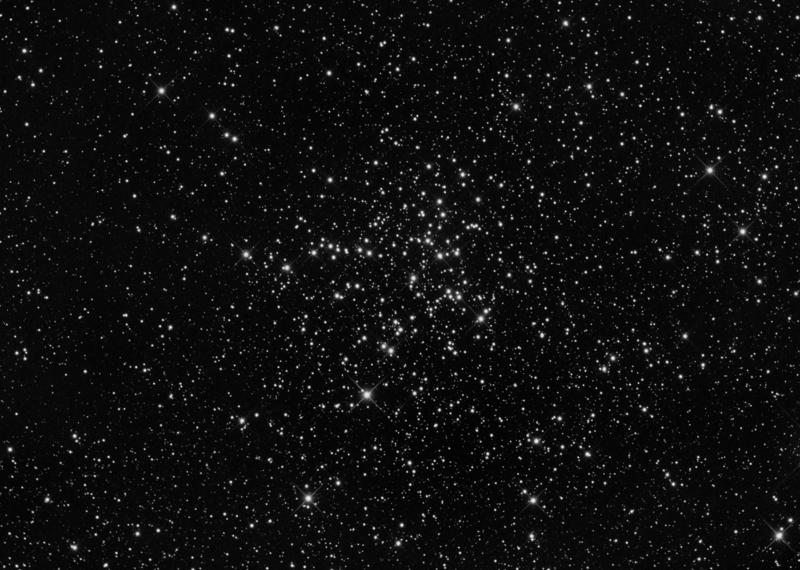}
		\caption{Astronomical \\ $(800 \times 570)$}
		\label{fig:train_astronomical}
		\vspace*{6pt}
	\end{subfigure}
	\begin{subfigure}{0.28\textwidth}
	    \captionsetup{justification=centering}
		\includegraphics[width=\textwidth]{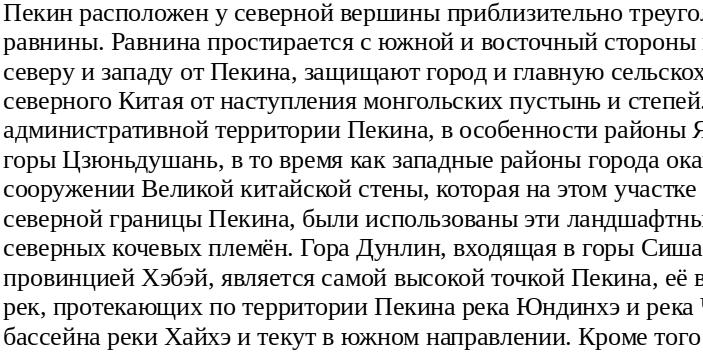}
		\caption{Text $(703 \times 354)$ \newline}
		\label{fig:train_text}
		\vspace*{6pt}
	\end{subfigure}
	\caption{Training images used for the external adaptation experiments (their actual sizes are included).}
	\label{fig:training}
\end{figure}

\begin{figure}
    \centering
	\begin{subfigure}{0.24\textwidth}
	    \captionsetup{justification=centering}
		\includegraphics[width=\textwidth]{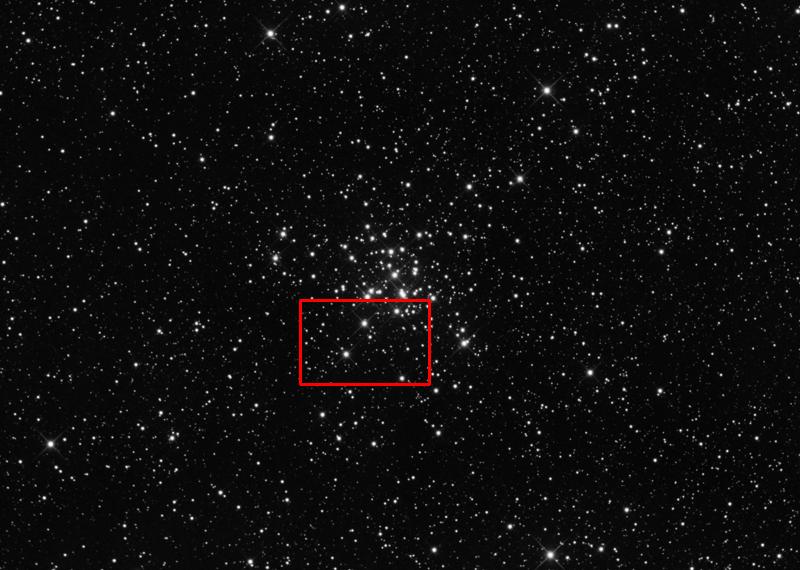}
		\caption{Clean astronomical \\ $(800 \times 570)$}
		\label{fig:clean_astronomical}
		\vspace*{6pt}
	\end{subfigure}
	\begin{subfigure}{0.24\textwidth}
	    \captionsetup{justification=centering}
		\includegraphics[width=\textwidth]{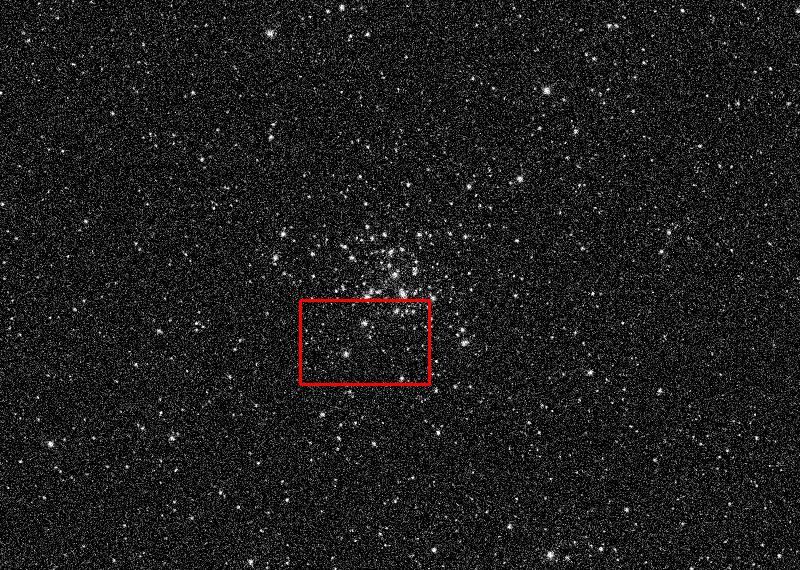}
		\caption{Noisy with $\sigma = 50$ \newline}
		\label{fig:noisy_astronomical}
		\vspace*{6pt}
	\end{subfigure}
	\begin{subfigure}{0.24\textwidth}
	    \captionsetup{justification=centering}
		\includegraphics[width=\textwidth]{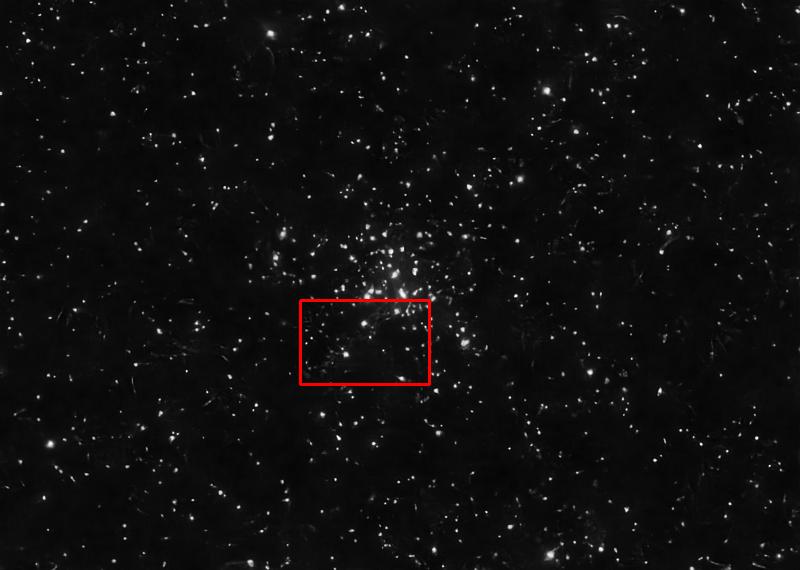}
		\caption{CDnCNN~\cite{Zhang_DnCnn_2017} \\ PSNR = 27.05dB}
		\label{fig:dncnn_astronomical}
		\vspace*{6pt}
	\end{subfigure}
	\begin{subfigure}{0.24\textwidth}
	    \captionsetup{justification=centering}
		\includegraphics[width=\textwidth]{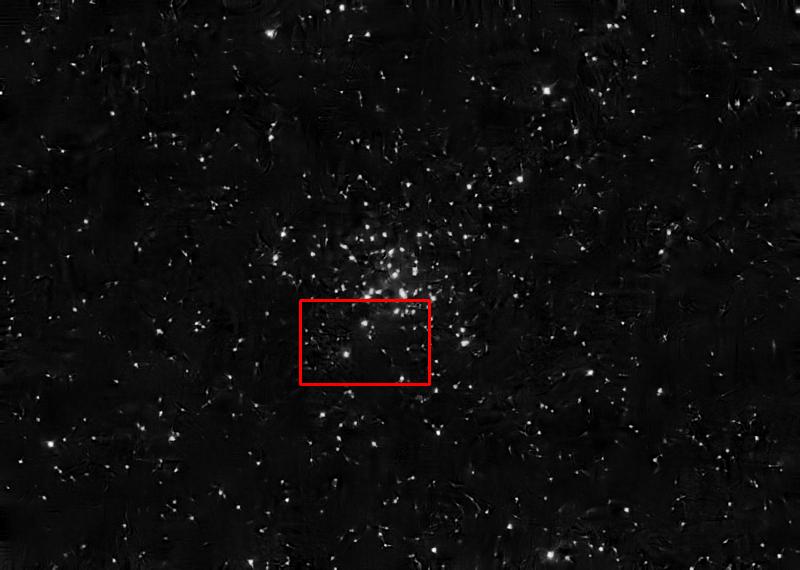}
		\caption{LIDIA (before adaptation) \\ PSNR = 26.44dB}
		\label{fig:nlms_d1_astronomical}
		\vspace*{6pt}
	\end{subfigure}
	\begin{subfigure}{0.23\textwidth}
	    \captionsetup{justification=centering}
		\includegraphics[width=\textwidth]{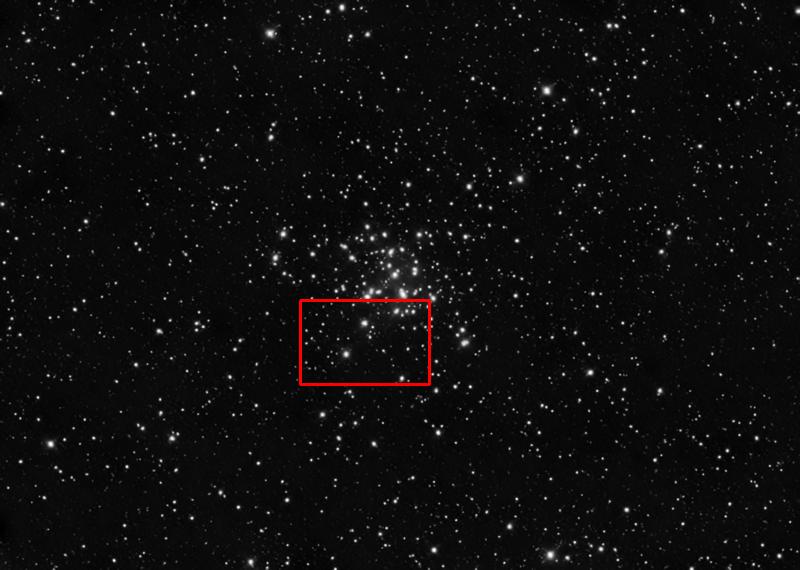}
		\caption{LIDIA (after adaptation) \\ PSNR = 28.04dB}
		\label{fig:nlms_d2_astronomical}
		\vspace*{6pt}
	\end{subfigure}
	\begin{subfigure}{0.25\textwidth}
	    \captionsetup{justification=centering}
		\includegraphics[width=\textwidth]{clean_m36_zoom}
		\caption{Clean \newline }
		\label{fig:clean_astronomical_zoom}
		\vspace*{6pt}
	\end{subfigure}
	\begin{subfigure}{0.24\textwidth}
	    \captionsetup{justification=centering}
		\includegraphics[width=\textwidth]{noisy_m36_zoom}
		\caption{Noisy \newline}
		\label{fig:noisy_astronomical_zoom}
		\vspace*{6pt}
	\end{subfigure}
	\begin{subfigure}{0.24\textwidth}
	    \captionsetup{justification=centering}
		\includegraphics[width=\textwidth]{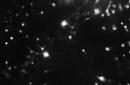}
		\caption{DnCNN \newline}
		\label{fig:dncnn_astronomical_zoom}
		\vspace*{6pt}
	\end{subfigure}
	\begin{subfigure}{0.24\textwidth}
	    \captionsetup{justification=centering}
		\includegraphics[width=\textwidth]{nlms_d1_m36_zoom}
		\caption{LIDIA (ours) \\ (before adaptation)}
		\label{fig:nlms_d1_astronomical_zoom}
		\vspace*{6pt}
	\end{subfigure}
	\begin{subfigure}{0.24\textwidth}
	    \captionsetup{justification=centering}
		\includegraphics[width=\textwidth]{nlms_d2_m36_zoom}
		\caption{LIDIA (ours) \\ (after adaptation)}
		\label{fig:nlms_d2_astronomical_zoom}
		\vspace*{6pt}
	\end{subfigure}
	\caption{An example of external adaptation for an astronomical image.}
	\label{fig:astronomical}
\end{figure}

\begin{figure}
    \centering
	\begin{subfigure}{0.24\textwidth}
	    \captionsetup{justification=centering}
		\includegraphics[width=\textwidth]{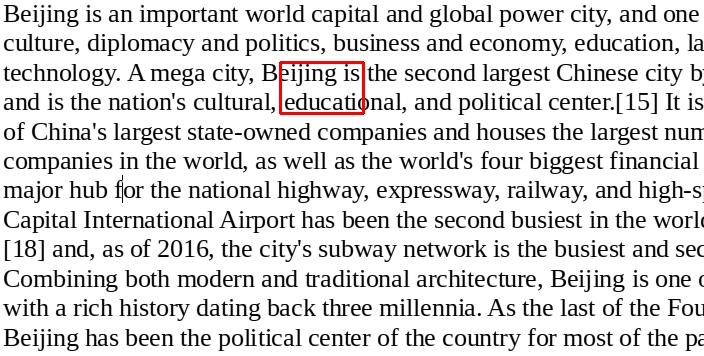}
		\caption{Clean text $(704 \times 356)$ \newline}
		\label{fig:clean_text}
		\vspace*{6pt}
	\end{subfigure}
	\begin{subfigure}{0.24\textwidth}
	    \captionsetup{justification=centering}
		\includegraphics[width=\textwidth]{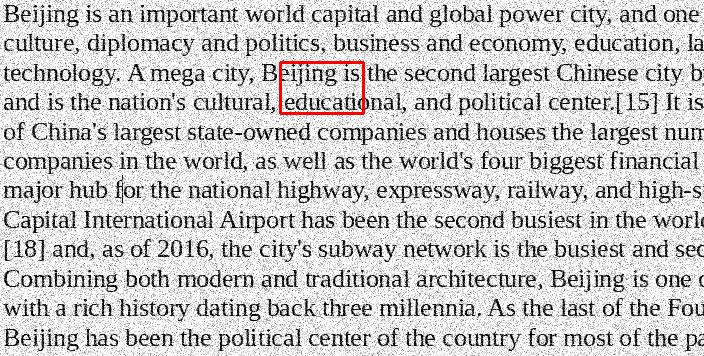}
		\caption{Noisy with $\sigma = 50$ \newline}
		\label{fig:noisy_tex}
		\vspace*{6pt}
	\end{subfigure}
	\begin{subfigure}{0.24\textwidth}
	    \captionsetup{justification=centering}
		\includegraphics[width=\textwidth]{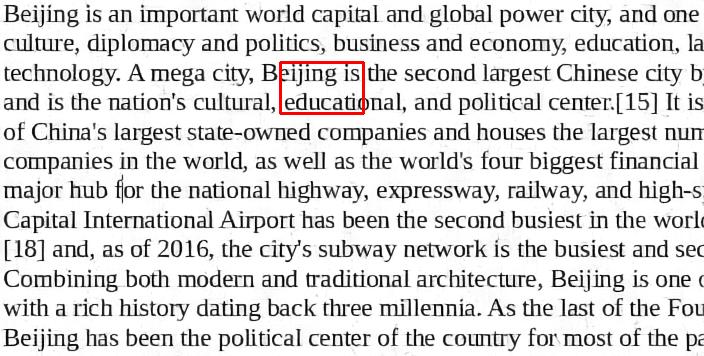}
		\caption{DnCNN~\cite{Zhang_DnCnn_2017} \\ PSNR = 23.33dB}
		\label{fig:dncnn_text}
		\vspace*{6pt}
	\end{subfigure}
	\begin{subfigure}{0.24\textwidth}
	    \captionsetup{justification=centering}
		\includegraphics[width=\textwidth]{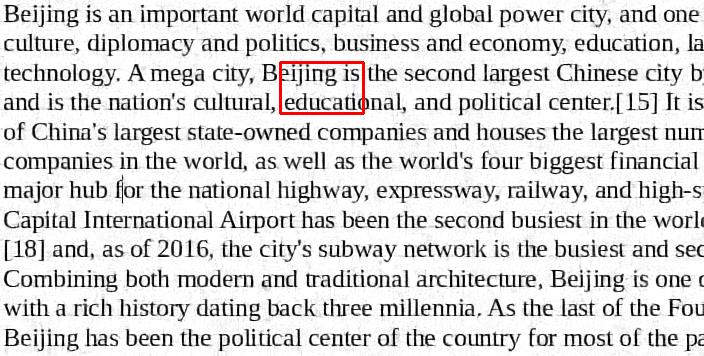}
		\caption{LIDIA (before adaptation) \\ PSNR = 22.52dB}
		\label{fig:nlms_d1_text}
		\vspace*{6pt}
	\end{subfigure}
	\begin{subfigure}{0.26\textwidth}
	    \captionsetup{justification=centering}
		\includegraphics[width=\textwidth]{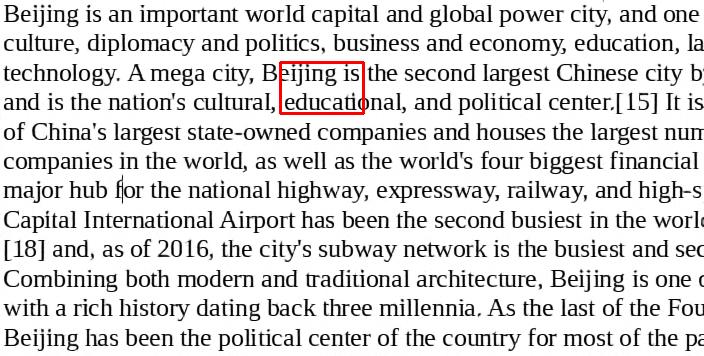}
		\caption{LIDIA (after adaptation) \\ PSNR = 26.78dB}
		\label{fig:nlms_d2_text}
		\vspace*{6pt}
	\end{subfigure}
	\begin{subfigure}{0.22\textwidth}
	    \captionsetup{justification=centering}
		\includegraphics[width=\textwidth]{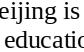}
		\caption{Clean text \newline}
		\label{fig:clean_text_zoom}
		\vspace*{6pt}
	\end{subfigure}
	\begin{subfigure}{0.24\textwidth}
	    \captionsetup{justification=centering}
		\includegraphics[width=\textwidth]{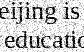}
		\caption{Noisy \newline }
		\label{fig:noisy_text_zoom}
		\vspace*{6pt}
	\end{subfigure}
	\begin{subfigure}{0.24\textwidth}
	    \captionsetup{justification=centering}
		\includegraphics[width=\textwidth]{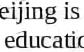}
		\caption{DnCNN \newline}
		\label{fig:dncnn_text_zoom}
		\vspace*{6pt}
	\end{subfigure}
	\begin{subfigure}{0.24\textwidth}
	    \captionsetup{justification=centering}
		\includegraphics[width=\textwidth]{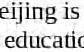}
		\caption{LIDIA (ours) \\ (before adaptation)}
		\label{fig:nlms_d1_text_zoom}
		\vspace*{6pt}
	\end{subfigure}
	\begin{subfigure}{0.24\textwidth}
	    \captionsetup{justification=centering}
		\includegraphics[width=\textwidth]{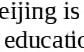}
		\caption{LIDIA (ours) \\ (after adaptation)}
		\label{fig:nlms_d2_text_zoom}
		\vspace*{6pt}
	\end{subfigure}
	\caption{An example of external adaptation for a text image. }
	\label{fig:text}
\end{figure}

\begin{figure}
    \centering
    \includegraphics[width=0.48\textwidth]{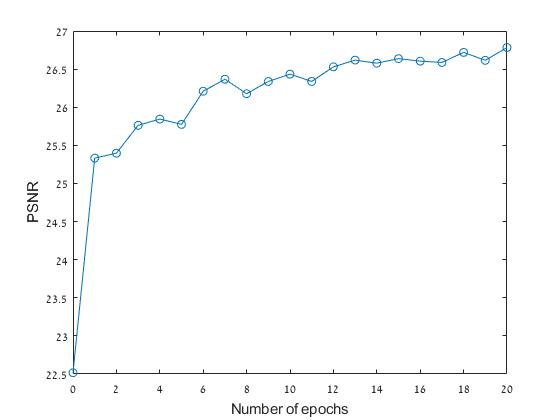}
    \caption{PSNR vs. number of epochs for the text image adaptation experiment.}
    \label{fig:psnr_vs_epochs}
\end{figure}

External adaptation is useful when the input image deviates from the statistics of the training images. Therefore, we demonstrate this adaptation on two non-natural images: an astronomical photograph and a scanned document. Applying a universal denoising machine on these images may create pronounced artifacts, as can be seen in Figures~\ref{fig:astronomical}~and~\ref{fig:text}. However, adapting LIDIA (our network), by training on a single similar image, significantly improves both the PSNR and the visual quality of the results. The training images used in the above experiments are shown in Figure~\ref{fig:training}. In the case of the text image, we used $20$ training epochs, due to very unique nature of its visual content. In accordance with the longer training time, this adaptation gains more than $4$dB in PSNR, where an improvement of $2.8$dB is achieved with only $3$ epochs, as illustrated in Figure~\ref{fig:psnr_vs_epochs}. 

\subsection{Internal Adaptation}

\begin{figure}
    \centering
	\begin{subfigure}{0.24\textwidth}
	    \captionsetup{justification=centering}
		\includegraphics[trim={25 0 25 0},clip, width=\textwidth]{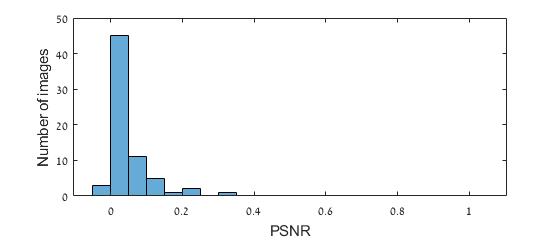}
		\caption{BSD68}
		\label{fig:histogram_bsd68_v2}
		\vspace*{6pt}
	\end{subfigure}
	\begin{subfigure}{0.24\textwidth}
	    \captionsetup{justification=centering}
		\includegraphics[trim={25 0 25 0},clip, width=\textwidth]{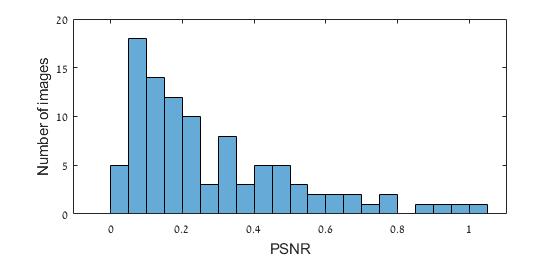}
		\caption{Urban100}
		\label{fig:histogram_urban100_v2}
		\vspace*{6pt}
	\end{subfigure}
	\caption{Histogram of improvement as a result of the internal adaptation applied on images from BSD68 and Urban100 sets.}
	\label{fig:histogram_bsd68_urban100_v2}
\end{figure}

\begin{figure}
    \centering
	\begin{subfigure}{0.235\textwidth}
	    \captionsetup{justification=centering}
		\includegraphics[width=\textwidth]{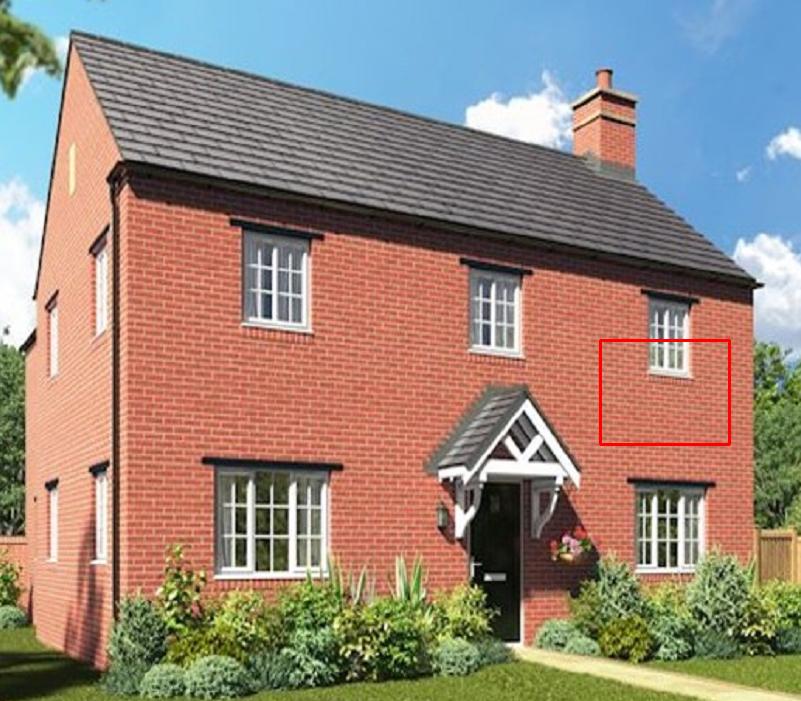}
		\caption{Clean \\ $(1024 \times 768)$}
		\label{fig:clean_brick_house}
		\vspace*{6pt}
	\end{subfigure}
	\begin{subfigure}{0.235\textwidth}
	    \captionsetup{justification=centering}
		\includegraphics[width=\textwidth]{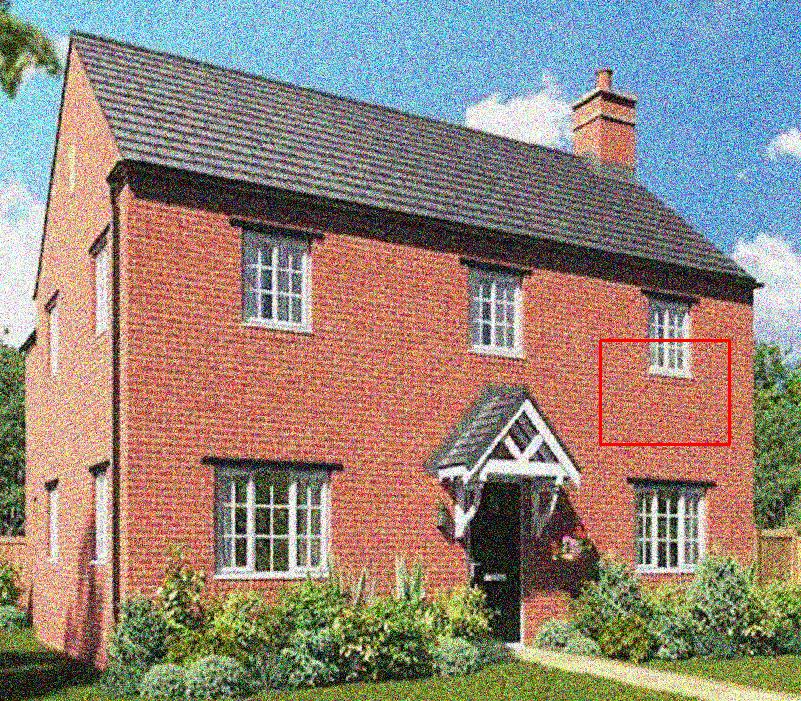}
		\caption{Noisy \\ with $\sigma = 50$}
		\label{fig:noisy_brick_house}
		\vspace*{6pt}
	\end{subfigure}
	\begin{subfigure}{0.235\textwidth}
	    \captionsetup{justification=centering}
		\includegraphics[width=\textwidth]{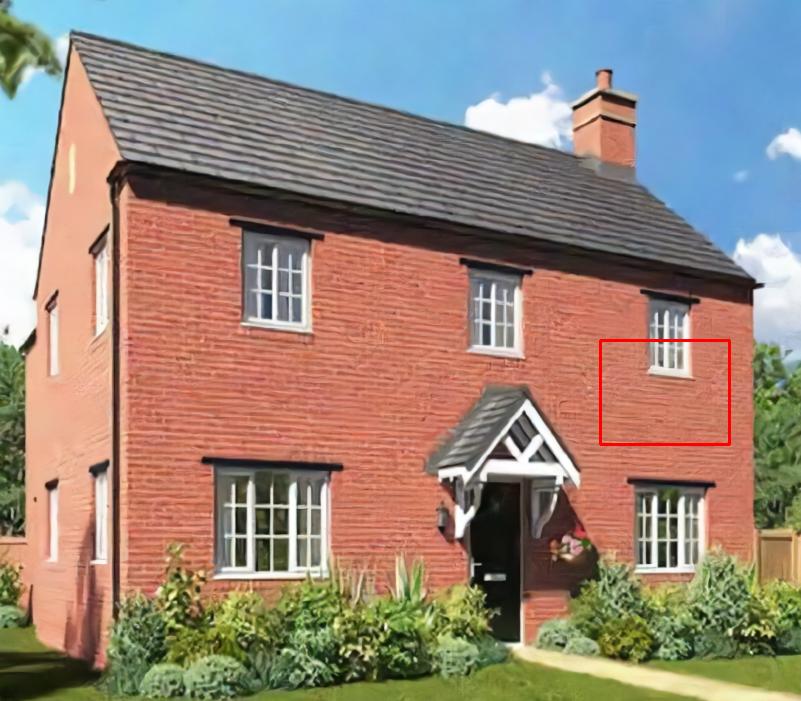}
		\caption{CDnCNN~\cite{Zhang_DnCnn_2017} \\ PSNR = 28.64dB}
		\label{fig:dncnn_brick_house}
		\vspace*{6pt}
	\end{subfigure}
	\begin{subfigure}{0.235\textwidth}
	    \captionsetup{justification=centering}
		\includegraphics[width=\textwidth]{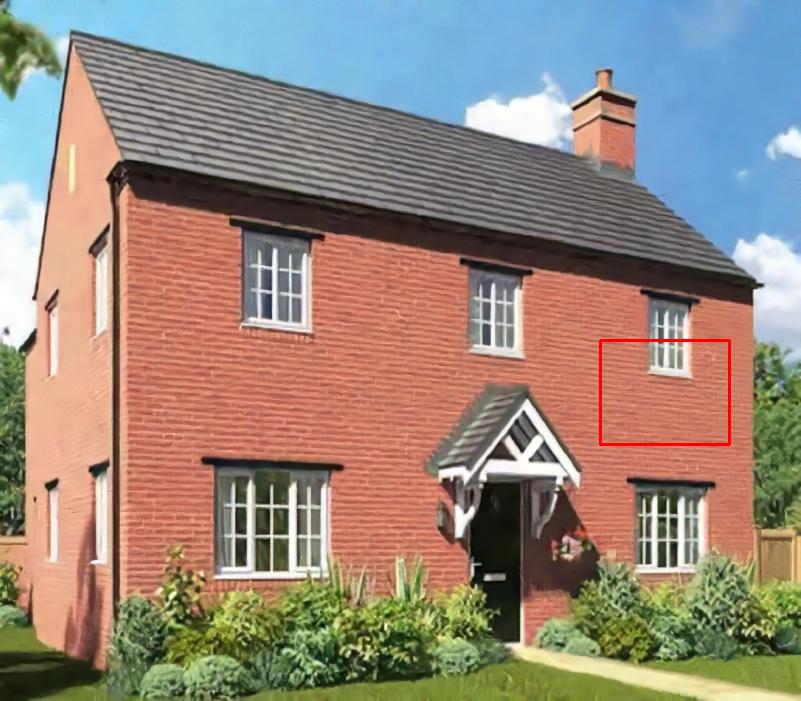}
		\caption{C-LIDIA (before adaptation)\\ PSNR = 28.77dB}
		\label{fig:nlms_d1_brick_house}
		\vspace*{6pt}
	\end{subfigure}
	\begin{subfigure}{0.225\textwidth}
	    \captionsetup{justification=centering}
		\includegraphics[width=\textwidth]{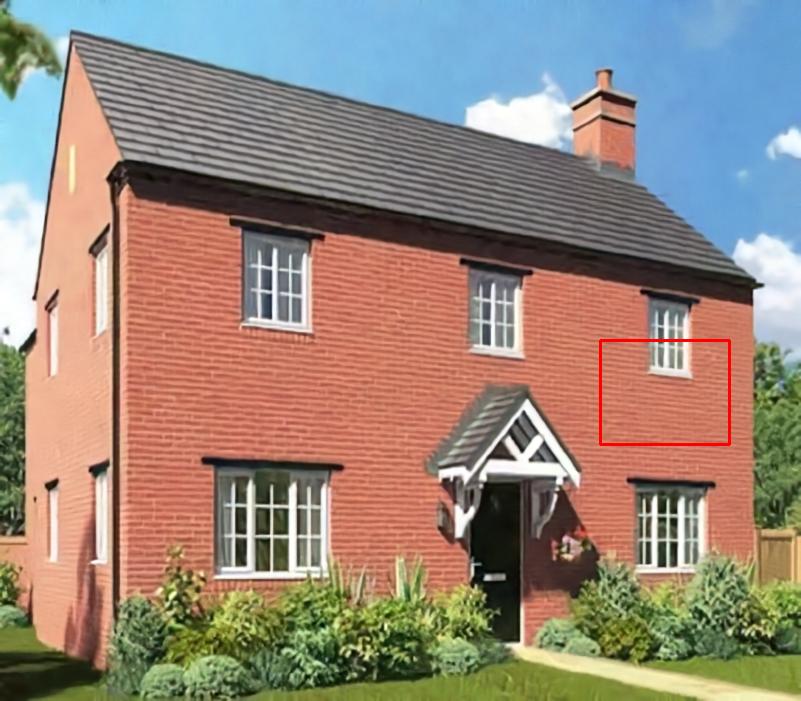}
		\caption{C-LIDIA (after adaptation) \\ PSNR = 29.28dB}
		\label{fig:nlms_d2_brick_house}
		\vspace*{6pt}
	\end{subfigure}
		\begin{subfigure}{0.245\textwidth}
	    \captionsetup{justification=centering}
		\includegraphics[width=\textwidth]{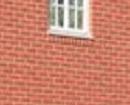}
		\caption{Clean \newline}
		\label{fig:clean_brick_house_zoom}
		\vspace*{6pt}
	\end{subfigure}
	\begin{subfigure}{0.235\textwidth}
	    \captionsetup{justification=centering}
		\includegraphics[width=\textwidth]{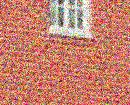}
		\caption{Noisy \newline }
		\label{fig:noisy_brick_house_zoom}
		\vspace*{6pt}
	\end{subfigure}
	\begin{subfigure}{0.235\textwidth}
	    \captionsetup{justification=centering}
		\includegraphics[width=\textwidth]{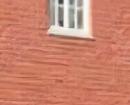}
		\caption{CDnCNN \newline}
		\label{fig:dncnn_brick_house_zoom}
		\vspace*{6pt}
	\end{subfigure}
	\begin{subfigure}{0.235\textwidth}
	    \captionsetup{justification=centering}
		\includegraphics[width=\textwidth]{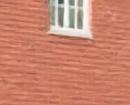}
		\caption{C-LIDIA (ours) \\ (before adaptation)}
		\label{fig:nlms_d1_brick_house_zoom}
		\vspace*{6pt}
	\end{subfigure}
	\begin{subfigure}{0.235\textwidth}
	    \captionsetup{justification=centering}
		\includegraphics[width=\textwidth]{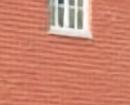}
		\caption{C-LIDIA (ours) \\ (after adaptation)}
		\label{fig:nlms_d2_brick_house_zoom}
		\vspace*{6pt}
	\end{subfigure}
	\caption{An example of internal adaptation.}
	\label{fig:brick_house}
\end{figure}

We proceed by reporting the performance of the internal adaptation method. Tables~\ref{tab:internal_adaptation_results}~and~\ref{tab:internal_adaptation_improvement} summarise quantitative denoising results of applying the proposed adaptation procedure on Urban100 and BSD68 image sets. These tables show that the internal adaptation can improve the denoising capability of the network, but the benefit may vary significantly from image to another, depending on its content. Figure~\ref{fig:histogram_bsd68_urban100_v2} presents histograms of  improvement per image for the two sets. As can be seen, the proposed adaptation has a minor effect on most of images in BSD68, but it succeeds rather well on Urban100, leading to a significant denoising boost. This difference can be attributed to the pronounced self similarity that is inherent in the Urban100 images. Figure~\ref{fig:brick_house} presents visual examples of the internal adaptation. Both images presented in these figures are characterised by a strong self similarity, which our adaptation is able to exploit.

As evident from Table~\ref{tab:internal_adaptation_improvement} and Figure~\ref{fig:histogram_bsd68_v2}, internal adaptation is not always successful, and it may lead to slightly decreased PSNR. However, we note that among the $168$ test images (BSD68 and Urban100) only two such failures were encountered, both leading to a degradation of $0.02$dB. 
Therefore, we conclude that internal adaptation is quite robust and cannot do much harm. Further work is required to identify a simple test for predicting the success of this adaptation, so as to invest the extra run-time only when appropriate. 

\begin{table}
    \centering
    \renewcommand{\arraystretch}{1.2}
    \setlength{\doublerulesep}{1pt}
    \begin{tabular}{|c||c|c|c|}
        \hline
        \multirow{2}{*}{Image set} & CDnCNN & C-LIDIA & C-LIDIA (ours) \\
        & \cite{Zhang_DnCnn_2017} & (ours) & with adaptation \\
        \hhline{|=#=|=|=|}
        Urban100 & 28.16 & 28.23 & 28.52 \\
        \hline
        BSD68 & 28.01 & 27.99 & 28.04 \\
        \hline
    \end{tabular}
    \caption{Performance of the internal adaptation on Urban100 and BSD68 sets for color images.}
    \label{tab:internal_adaptation_results}
\end{table}

\begin{table}
    \centering
    \renewcommand{\arraystretch}{1.2}
    \setlength{\doublerulesep}{1pt}
    \begin{tabular}{|c||c|c|c|c|}
        \hline
        Image set & Maximum & Minimum & Average & Median \\
        \hhline{|=#=|=|=|=|}
        Urban100 & 1.09 & 0.01 & 0.29 & 0.21 \\
        \hline
        BSD68 & 0.35 & -0.02 & 0.05 & 0.04 \\
        \hline
    \end{tabular}
    \caption{PSNR improvement obtained as a result of applying internal adaptation.}
    \label{tab:internal_adaptation_improvement}
\end{table}

\section{Conclusion}
\label{sec:conclusion}

This work starts by presenting a lightweight universal network for supervised image denoising. Our patch-based architecture exploits non-local self-similarity and representation sparsity, augmented by a multiscale treatment. Separable linear layers, combined with non-local $k$ neighbor search, allow capturing non-local interrelations between pixels using small number of learned parameters. The proposed network achieves SOTA results in the lightweight category, and competitive performance overall. On top of the above, this work offers two image-adaption techniques, external and internal, both aiming for improved denoising performance by better tuning the above universal network to the incoming noisy image. We demonstrate the effectiveness of these methods on images with unique content or having significant self-similarity. 


		
\bibliography{bibliography_file}{}
\bibliographystyle{IEEEtran}
\end{document}